\documentclass[12pt]{article}
\pdfoutput=1
\usepackage{amsmath,amssymb,amsfonts,graphicx,slashed,color,cite}
\usepackage{color}
\usepackage[colorlinks=true,linkcolor=blue,citecolor=blue,linktoc=all]{hyperref}

\def\uiucaddress{\small Department of Physics, University of Illinois, 1110 W. Green St., 
Urbana IL 61801-3080, U.S.A. }
\def\title{\LARGE{Shape Dependence of Entanglement Entropy in Conformal Field Theories
}}

\parskip=\baselineskip

\newcommand{\myfig}[3]{
	\begin{figure}[ht]
	\centering
	\includegraphics[width=#2cm]{#1}\caption{#3}\label{fig:#1}
	\end{figure}
	}

\newcommand{\cDsl}{{{\cal D}\kern-.65em /\,}}
\newcommand{\cDslsm}{{{\cal D}\kern-.5em /\,}}
\newcommand{\nabslsm}{\nabla\kern-.55em /}
\newcommand{\pasl}{\pa\kern-.55em /}
\newcommand{\psl}{p\kern-.45em /}
\newcommand{\Dsl}{D\kern-.65em /}
\newcommand{\Asl}{A\kern-.55em /}
\newcommand{\nabsl}{\nabla\kern-.65em /\kern+.2em}
\newcommand{\qsl}{q\kern-.5em /}
\newcommand{\ksl}{k\kern-.5em /}
\newcommand{\rsl}{r\kern-.5em /}

\newcommand{\cDslLCsq}{{\stackrel{\circ}{\cDsl^{\kern2pt 2}}}}

\newcommand\cc[1]{#1^{^{\kern-6pt \circ}}\kern2pt}

\newcommand{\re}{\mathbb{R}}

\newcommand{\bx}{\boldsymbol{x}}

\newcommand{\pa}{\partial}

\newcommand{\beq}{\begin{equation}}
\newcommand{\eeq}{\end{equation}}
\newcommand{\beqn}{\begin{eqnarray}}
\newcommand{\eeqn}{\end{eqnarray}}

\newcommand{\ui}{\overline{i}}
\newcommand{\uj}{\overline{j}}
\newcommand{\bdx}{\mathbf{x}}

\newcommand{\brho}{\widehat{\rho}}
\newcommand{\hT}{\widehat{T}}
\newcommand{\hK}{\widehat{K}}

\newcommand{\n}{\chi}

\def\dalemb#1#2{{\vbox{\hrule height .#2pt
\hbox{\vrule width.#2pt height#1pt \kern#1pt
\vrule width.#2pt}
\hrule height.#2pt}}}

\begin{document}

\begin{center}
\title
\end{center}
\vskip 2 cm
\centerline{{\bf 
Thomas Faulkner, Robert G. Leigh and Onkar Parrikar}}

\vspace{.5cm}
\centerline{\it \uiucaddress}
\vspace{2cm}

\begin{abstract}
\noindent We study universal features in the shape dependence of entanglement entropy in the vacuum state of a conformal field theory (CFT) on $\re^{1,d-1}$. We consider the entanglement entropy across a deformed planar or spherical entangling surface in terms of a perturbative expansion in the infinitesimal shape deformation. In particular, we focus on the second order term in this expansion, known as the entanglement density. This quantity is known to be non-positive by the strong-subadditivity property. We show from a purely field theory calculation that the non-local part of the entanglement density in any CFT is universal, and proportional to the coefficient $C_T$ appearing in the two-point function of stress tensors in that CFT. As applications of our result, we prove the conjectured universality of the corner term coefficient $\frac{\sigma}{C_T}=\frac{\pi^2}{24}$ in $d=3$ CFTs, and the holographic Mezei formula for entanglement entropy across deformed spheres.
\end{abstract}

\pagebreak
%
%
\section{Introduction}
Entanglement entropy has  a central position in the study of quantum field theories. It is a powerful tool to probe the structure of quantum states, primarily because: (i) it is sufficiently non-local to capture certain global properties, and (ii) it is geometric by definition and hence universal in its applicability. As a result, entanglement entropy has provided great insights in a wide class of systems such as relativistic field theories \cite{Calabrese:2004eu,1751-8121-42-50-504005,Casini:2009sr}, conformal field theories (CFTs) \cite{Casini:2011kv, Faulkner:2013yia, Faulkner:2014jva}, topologically ordered phases of matter \cite{Kitaev:2005dm,PhysRevLett.96.110405,Dong:2008ft,lihaldane}, strongly-coupled theories with holographic duals \cite{Ryu:2006bv,Ryu:2006ef,Hubeny:2007xt}, etc. It has also become clear that entanglement will play a crucial role in understanding the emergence of geometry in the AdS/CFT correspondence \cite{VanRaamsdonk:2010pw,Faulkner:2013ica}. Despite this, computing entanglement entropy for arbitrary shaped regions in general dimension still remains a non-trivial task, especially outside the arena of quantum field theories with classical gravitational duals. While much progress can be made in special symmetric cases such as the entanglement entropy across planar surfaces in relativistic quantum field theories, spherical surfaces in CFTs, etc., it is desirable to develop a larger theoretical toolkit.
\myfig{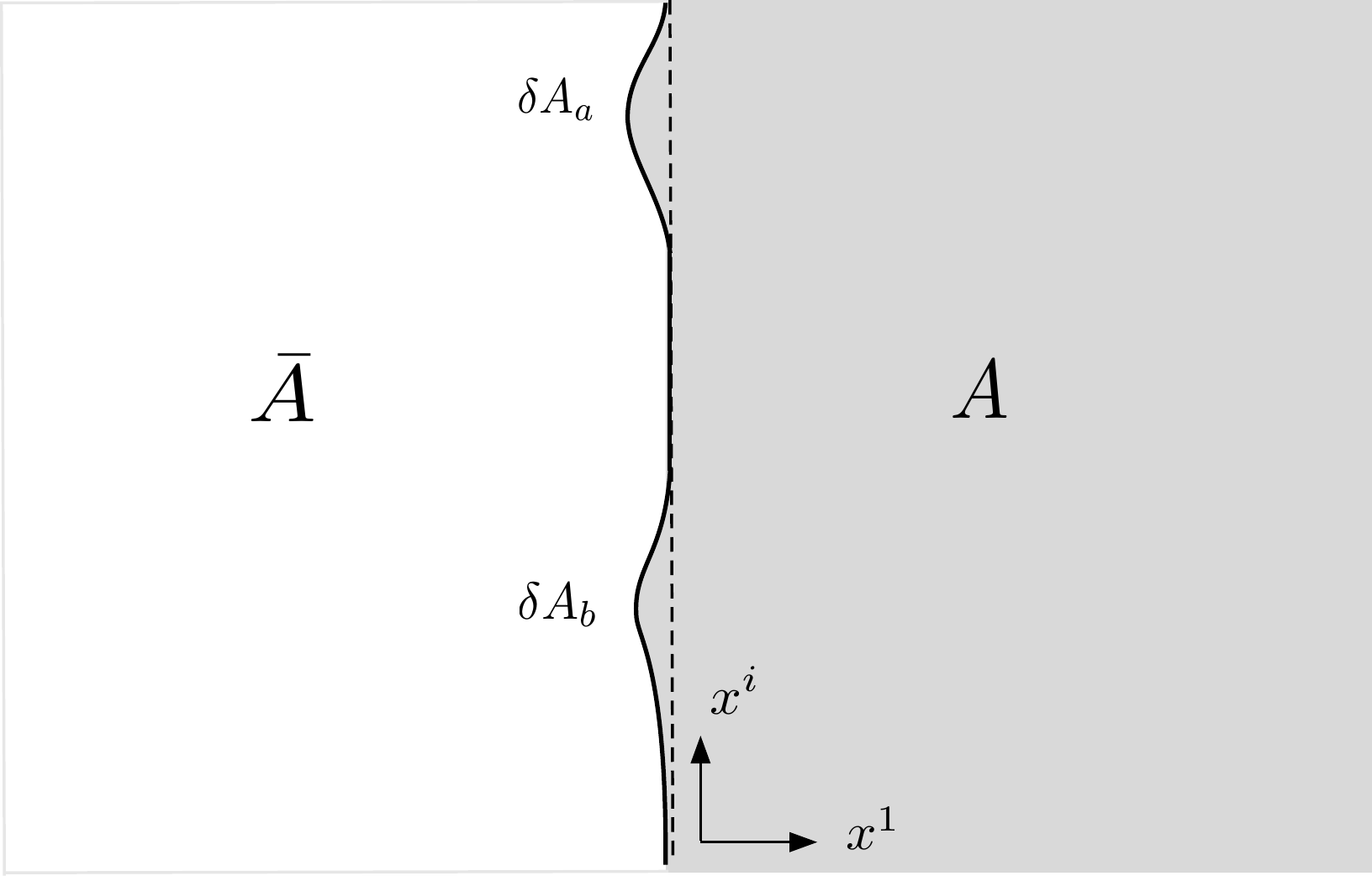}{8}{\textsf{The set-up for the planar case: the original subregion $A$ is the half-space $x^1>0$, with the entangling surface $x^1=0$ (dashed line). We deform the entangling surface to $x^1=-\epsilon\;\chi(x^i)$ (bold line) by glueing on the area elements $\delta A_{a,b}$ at points $x_{a,b}$ along the entangling surface.}}

In this paper, we study entanglement entropy for \emph{deformed} half-spaces and ball-shaped regions in the vacuum state of a conformal field theory on $\re^{1,d-1}$. To be concrete, we first explain our construction for deformed half-spaces (see Fig.~\ref{fig:fig1}). Let us pick global coordinates $\bx^{\mu}=(x^0, x^1,\cdots, x^{d-1})$ on $\re^{1,d-1}$, where $x^0$ is the time coordinate. Pick the Cauchy surface $x^0=0$, and consider the reduced density matrix $\brho_0$ on the half space $A$ given by
\beq
A=\left\{\bx^{\mu}\in \re^{1,d-1} | x^0=0,\;x^1 > 0\right\}.
\eeq
The entanglement entropy between $A$ and its complement $\bar{A}$ is defined as the Von Neumann entropy of $\brho_0$. Next, deform the region $A$ slightly to
\beq
A + \delta A = \left\{\bx^{\mu}\in \re^{1,d-1}\ \Big|\ x^0=0,\;x^1 > - \epsilon\; \n(x^2,\cdots, x^{d-1})\right\}
\eeq
where $\n(x^2,\cdots, x^{d-1})$ is a smooth function of the $(d-2)$ transverse spatial coordinates (parametrizing the entangling surface) which we denote collectively as $x^i=(x^2,\cdots,x^{d-1})$, and $\epsilon$ is a positive infinitesimal parameter. This corresponds to deforming the entangling surface to $x^1=-\epsilon\chi(x^i)$ within the original Cauchy surface. We can also generalize this and deform the entangling surface by the infinitesimal vector field $\zeta^{\ui}(x^i)$, which we take to lie in the plane perpendicular to the original surface $x^1=0$ (i.e., the overlined indices run over $\ui = 0,1$), and which also includes, for instance, time-like deformations. The entanglement entropy across the deformed surface can be written as a perturbative expansion in $\zeta^{\ui}$
\beq
S_{EE}[A+\delta A] = S_{EE}[A]+\int d^{d-2}x\;\zeta^{\ui}(x)S^{(1)}_{\ui}(x)+\frac{1}{2!}\int d^{d-2}x_a d^{d-2}x_b\;\zeta^{\ui}(x_a)\zeta^{\uj}(x_b)\;S^{(2)}_{\ui\uj}(x_a,x_b)+\cdots
\eeq
The quantity $S^{(2)}_{\ui\uj}(x_a,x_b)$ is known as the \emph{entanglement density} \cite{Nozaki:2013wia,Nozaki:2013vta,Bhattacharya:2014vja}\footnote{Sometimes entanglement density
is defined with an extra minus sign to make it a naturally positive quantity, see \eqref{neg.}. 
It is also not clear why one should think of it as a density - \emph{entanglement susceptibility} would
probably be a more appropriate name; however we will follow \cite{Nozaki:2013wia,Nozaki:2013vta,Bhattacharya:2014vja} in using the term entanglement density. A similar quantity
was studied in \cite{Bousso:2015mna}.}, and will be the primary focus of the present paper. In the case where $\zeta^{\ui}$ is spacelike and $\zeta^{1}>0$, there is another nice way of thinking about the entanglement density: start with the half space $A$ and glue on to it small area elements $\delta_a A^{\ui}$ and $\delta_b A^{\ui}$ at the points $x_a$ and $x_b$ on the entangling surface
such that $\delta_a A$ and $\delta_b A$ are non overlapping. Then to lowest order in $\delta_{a,b}A$, the entanglement density is proportional to the \emph{conditional mutual information} between $\delta A_a$ and $\delta A_b$ given the state on $A$ 
\beqn \label{cmi}
\delta_a A^{\ui}\;\delta_b A^{\uj}\; S^{(2)}_{\ui\uj}(x_a,x_b) &=& S_{EE}[A+\delta_a A+\delta_bA]-S_{EE}[A+\delta_aA]-S_{EE}[A+\delta_bA]+S_{EE}[A]\nonumber\\
&=& -I(A+\delta A_a; \delta A_b)+ I(A;\delta A_b)\nonumber\\
&=& -I(\delta A_a; \delta A_b|A)
\eeqn
where $I(X;Y)$ is the mutual information between the regions $X$ and $Y$. The strong subadditivity property then implies
\beq\label{neg.}
\delta_a A^{\ui}\;\delta_b A^{\uj}\; S^{(2)}_{\ui\uj} \leq 0.
\eeq
Consequently, the entanglement density provides a natural notion of a metric on the space of geometries of the entangling surface. In theories with holographic duals, the Ryu-Takayanagi proposal maps this space into the space of mimimal-area surfaces in the bulk, and so the entanglement density provides a natural metric on the latter space as well  (see \cite{Czech:2015qta} for more details in the $AdS_3$/CFT$_2$ case). It has also been argued in \cite{Bhattacharya:2014vja} that in holographic theories, equation \eqref{neg.} applied to a special class of deformations maps to the integrated null-energy condition on the bulk minimal-area surface. 

In general, entanglement density in conformal field theories can contain two types of terms: (1) contact terms which arise in the coincident limit $x_a \to x_b$, and (2) a non-local term which is finite and well-defined when $x_a$ and $x_b$ are separated. For the most part, we will be interested in the latter.
This non-local term is isolated via the definition \eqref{cmi} in terms of the conditional mutual information
which makes it is clear this term should be independent of the UV cutoff. 
The main result of the present paper is as follows: \emph{for any conformal field theory, the non-local term in the entanglement density for a planar surface is universal and given by}
\beq\label{planes}
S^{(2)}_{\ui\uj,non-local}(x_a,x_b) = -\frac{2\pi^2 C_T}{(d+1)}\frac{\eta_{\ui\uj}}{|x_a-x_b|^{2(d-1)}} 
\eeq
\emph{where $C_T$ is the numerical coefficient appearing in the two-point function of stress tensors in the CFT.} Equation \eqref{planes} was obtained in \cite{Nozaki:2013vta} for a class of holographic theories using the Ryu-Takayanagi formula. However, we emphasize that in this paper we are working with completely general CFTs.\footnote{Actually it is enough to invoke the $SO(1,d-1) \times SO(1,1)$ conformal symmetries that leave the entangling surface $\partial A$ fixed, including the boosts in the transverse plane, to argue that the cut-off independent part of the entanglement density should take the form as in \eqref{planes}. Here we will be tracking down the overall coefficient.}
 We will employ purely field theoretic techniques (developed in \cite{Rosenhaus:2014woa, Rosenhaus:2014zza, Allais:2014ata, Faulkner:2014jva}) to prove equation \eqref{planes}, thus extending the validity of this formula to arbitrary conformal field theories with or without holographic duals, and further providing a non-trivial check on the Ryu-Takayanagi and the Hubeny-Rangamani-Takayanagi proposals for entanglement entropy in holographic theories.

\myfig{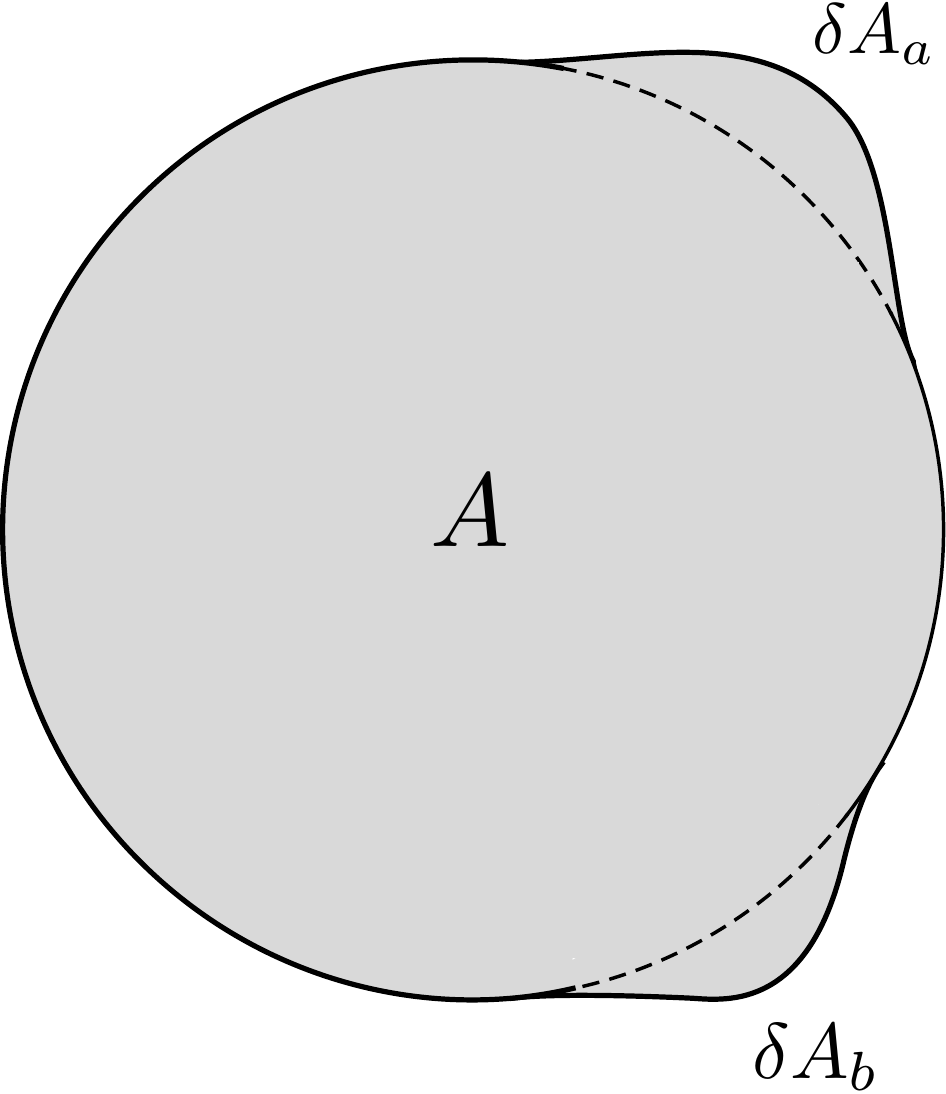}{4.5}{\textsf{The set-up for the spherical case. We deform the original entangling surface $\bdx^2=R^2$ (dashed line) by glueing on the area elements $\delta A_{a,b}$ at points $\Omega_{a,b}$ along the entangling surface.}}
An analogous formula also holds in the case where we take $A$ to be a ball-shaped region of radius $R$ (see Fig.~\ref{fig:fig2}). Let $\bdx=(x^1,\cdots, x^{d-1})$ denote the spatial coordinates on the Cauchy slice $x^0=0$ and take $A$ to be the region $\bdx^2 \leq R^2$. For $\bdx_a = R\;\Omega_a$ and $\bdx_b = R\;\Omega_b$ two well separated points on the entangling surface, the non-local term in the entanglement density is given by
\beq\label{spheres}
S^{(2)}_{\ui\uj,non-local}(\Omega_a,\Omega_b) =- \frac{2\pi^2 C_T}{(d+1)R^{2}}\frac{\eta_{\ui\uj}}{|\Omega_a-\Omega_b|^{2(d-1)}}. 
\eeq
In fact, since a ball-shaped region can be mapped into a half-space by a conformal tranformation, we will argue that equation \eqref{spheres} follows as a direct consequence of equation \eqref{planes} in a CFT.  

A number of results follow as corollaries: (i) in \cite{Bueno:2015rda, Bueno:2015xda}, it was conjectured based on holographic and numerical evidence that the coefficient $a(\theta)$ of the corner term contribution to the entanglement entropy in $d=3$ CFTs has the universal behaviour 
\beq
\lim_{\theta \to \pi} a(\theta) =\sigma(\pi - \theta)^2+\cdots, \qquad
\frac{\sigma}{C_T}=\frac{\pi^2}{24}
\eeq 
where $\theta$ is the opening angle of the corner. We will prove this conjecture as a special case of our results. (ii) We also prove the Mezei formula for the universal part of the entanglement entropy across deformed spheres
\beq
\label{mezei}
S^{(2)}_{EE}= C_T\frac{\pi^{\frac{d+2}{2}}(d-1)}{2^{d-2}\Gamma(d+2)\Gamma(d/2)}\sum_{\ell,m_1,\cdots,m_{d-3}}a^2_{\ell,m_1,\cdots,m_{d-3}}\prod_{k=1}^d(\ell+k-2)\times \begin{cases} (-1)^{\frac{d-1}{2}}\frac{\pi}{2} & d\;\mathrm{odd} \\ (-1)^{\frac{d-2}{2}}\mathrm{ln}\,\frac{R}{\delta} & d\;\mathrm{even} \end{cases}
\eeq
which was conjectured in \cite{Mezei:2014zla} based on holographic calculations in a large class of theories. In \eqref{mezei}, $a_{\ell,m_1,\cdots,m_{d-3}}$ are the coefficients of the expansion of the shape deformation in terms of real hyperspherical harmonics on the entangling surface. The Mezei formula is meant to apply to the universal term in CFT entanglement entropy
for a deformed sphere, and the \emph{positivity} of the overall coefficient demonstrates that the sphere locally minimizes this universal term in the space of shapes, suggesting that the sphere is somehow the optimal measure of degrees of freedom in a CFT for use as an RG monotone. Further, the above formula was used in \cite{Bueno:2015lza} to compute universal corner contributions to entanglement entropy in higher dimensions. Therefore, our proof of the Mezei formula also completes the proof of universality of corner contributions in higher dimensions. In this way, our CFT calculation fits nicely into the triangle of recent studies and conjectures \cite{Nozaki:2013vta, Allais:2014ata, Mezei:2014zla, Bueno:2015lza, Bueno:2015qya, Bueno:2015rda, Bueno:2015xda, Miao:2015dua, Bueno:2015ofa} (see also \cite{Klebanov:2012yf, Huang:2015bna, Carmi:2015dla, Elvang:2015jpa, Fonda:2015nma}) on entanglement density, corner contributions and the Mezei formula.

The rest of the paper is organized as follows: in section \ref{prelim}, we review some elementary facts about entanglement across planar and spherical surfaces, which will be relevant for our subsequent calculations. In section \ref{comp}, we present the CFT calculation of the universal non-local term in the entanglement density for planar and spherical surfaces. In section \ref{app}, we will then use our result for the entanglement density to prove the universality of corner contributions in $d=3$ CFTs and the Mezei formula. Finally, we will end with some discussion about prospects for future work.

\section{Preliminaries}\label{prelim}
Entanglement entropy is defined as follows -- consider the density matrix $|\Psi\rangle\langle\Psi|$ corresponding to a pure state defined on the Cauchy surface $\Sigma$. In this paper, we will take $|\Psi\rangle$ to be the ground state of a conformal field theory. Let us partition $\Sigma$ into two subregions $A$ and $\bar{A}$. For local quantum field theories, we expect the Hilbert space $\mathfrak{h}_{\Sigma}$ to factorize into the tensor product $\mathfrak{h}_{\Sigma}=\mathfrak{h}_{A}\otimes \mathfrak{h}_{\bar{A}}$. If this is the case, we can trace over $\mathfrak{h}_{\bar{A}}$ to obtain the reduced density matrix
\beq
\brho_0 = \mathrm{tr}_{\mathfrak{h}_{\bar{A}}}(|\Psi\rangle\langle\Psi|)
\eeq 
which contains all the relevant information pertaining to the subregion $A$. Then the \emph{entanglement entropy}  between $A$ and $\bar{A}$ is defined as the von Neumann entropy of $\brho_0$
\beq
S_{EE}[A] = -\mathrm{tr}_{\mathfrak{h}_A}\left(\brho_0\,\mathrm{ln}\;\brho_0\right).
\eeq
In this context, the boundary $\pa A$ of $A$ is referred to as the \emph{entangling surface}. It is also useful to define the \emph{modular Hamiltonian} (also known as the entanglement Hamiltonian) $\widehat{H}_E$ in terms of $\brho_0$ as
\beq
\brho_0 \equiv e^{-\widehat{H}_E}.
\eeq
In general, the modular Hamiltonian is not a local operator, in the sense that the modular evolution $U(s)=\brho_0^{\;is}$ does not map local operators to local operators. However, there are a few special cases where symmetry forces the modular Hamiltonian to be local. The simplest such case is when we take $A$ to be the half-space $x^1>0$. In this case, the modular Hamiltonian takes a simple form in terms of the CFT defined on Euclidean space $\re^d$: 
\beq\label{mh}
\widehat{H}_{E,plane} = 2\pi\widehat{K}+ \mathrm{constant}
\eeq
where $\widehat{K}$ is the generator of rotations around the entangling surface in the $(x^0_E, x^1)$ plane ($x^0_E$ is Euclidean time)
\beq
\widehat{K}=\int d^{d-2}x^i\int_0^{\infty}dx^1\;x^1\;\widehat{T}^{00}(0,x^1,x^i) 
\eeq
and the constant in \eqref{mh} is chosen such that $\mathrm{tr}_{\mathfrak{h}_A}\brho_0 =1$. This is known as the Bisognano-Wichmann theorem \cite{Bisognano:1976za}. The fact that the modular Hamiltonian for planar entangling surfaces in the vacuum state of a conformal field theory on $\re^{1,d-1}$ is local, and can be written as an integral over the stress tensor will play a crucial role in our calculation of the entanglement density. In fact, the statement of the Bisognano-Wichmann theorem is true for the vacuum state of any relativistic quantum field theory, irrespective of conformal symmetry, and so it should be possible to extend our calculation to the more general class of relativistic quantum field theories. However, in this paper we will restrict ourselves to CFTs, because the calculation simplifies greatly in this case. 

In conformal field theories, the modular Hamiltonian for a ball-shaped region (of radius $R$) is also local \cite{Casini:2011kv}. This happens because the conformal transformation 
\beq
\psi^{\mu}(\bx)=\frac{\bx^{\mu}-(\bx\cdot \bx)C^{\mu}}{1-2(C\cdot \bx)+(\bx\cdot \bx)(C\cdot C)}+2R^2C^{\mu}
\eeq
with $C^{\mu} = (0,\frac{1}{2R},0,\cdots, 0)$, maps the half-space $x^1>0$ to the ball-shaped region $\bdx^2 \leq R^2$. Since conformal transformations are symmetries in a CFT, such a map leaves the ground state invariant, and  the reduced density matrix on the ball-shaped region can be related to the reduced density matrix on the half-space by a unitary transformation. Additionally, one can transplant the modular Hamiltonian from the half-space to the ball-shaped region by pushing forward the modular flow by $\psi$, which gives
\beq
\widehat{H}_{E,sphere} = 2\pi \int_{\bdx^2\leq R^2} d^{d-1}\bdx\;\frac{R^2-\bdx^2}{2R}\hT^{00}(\bdx)+\mathrm{constant}'.
\eeq
For this reason, the calculation of the entanglement density for ball-shaped regions is no more difficult than the calculation for half-spaces in CFTs. 

\section{The CFT computation}\label{comp}
Let us now delve into the calculation of the entanglement density in conformal field theories. For simplicity, we will describe the computation for half spaces in some detail, and then derive the corresponding result for ball-shaped regions by using the conformal transformation mentioned previously. So take $A$ to be the half-space $x^1>0$. Consider now the entanglement entropy of the deformed region $A+\delta A$ given by $x^1>-\epsilon\chi(x^i)$. In order to compute this entropy, we can use the coordinate transformation 
\beq\label{deformation}
\bx^{\mu}\to \bx^{\mu}-\zeta^{\mu}(x),\;\;\;\zeta^{\mu}=-\left(0,\epsilon\n(x^i),0,\cdots,0\right)
\eeq
to map the deformed entangling region $A+\delta A$ back to the half-space $A$. However, we must bear in mind that such a coordinate transformation has a non-trivial action on the metric. In terms of the new coordinates, the metric is given by
\beq\label{defmetric}
g_{\mu\nu}=\eta_{\mu\nu}+2\pa_{(\mu}\zeta_{\nu)}+O(\epsilon^2).
\eeq
Therefore, to compute the entanglement entropy for $A+\delta A$ in flat space, we may equivalently compute the entanglement entropy for the half-space $A$ but with the deformed metric $g_{\mu\nu}$ \cite{Banerjee:2011mg}\footnote{
Note that \eqref{eediff} is true (even for the UV divergent terms) if we use a ``covariant'' regulator to define EE \cite{Casini:2015woa,Liu:2012eea,Grover:2011fa}. However since we are ultimately interested in a UV finite quantity the regulator used at intermediate stages in the calculation should not matter. }
\beq \label{eediff}
S_{EE}[A+\delta A,\eta_{\mu\nu}]=S_{EE}[A,g_{\mu\nu}].
\eeq
For our purpose it suffices to keep only the term linear in $\zeta^{\mu}$ in equation \eqref{defmetric} because we are interested in computing the non-local contribution to the entanglement entropy at second order in the perturbation series, while the $O(\epsilon^2)$ terms in \eqref{defmetric} can at most generate a local contribution at this order. The shape deformation in \eqref{deformation} is somewhat special in that it preserves the Cauchy surface $x^0=0$. In our calculation we will relax this and consider the more general deformation
\beq\label{deformation2}
\zeta = \zeta^{\ui}(x^i)\pa_{\ui}=\zeta^0(x^i)\pa_0+\zeta^1(x^i)\pa_1
\eeq
which also includes time-like deformations of the entangling surface, and of which equation \eqref{deformation} is a special case.\footnote{We need not include components along the transverse directions $\pa_i$ because these simply amount to reparametrizations of the entangling surface, which do not change the entanglement entropy.} 

The advantage of trading the original problem of computing $S_{EE}[A+\delta A, \eta_{\mu\nu}]$ with that of computing $S_{EE}[A,\eta_{\mu\nu}+2\pa_{(\mu}\zeta_{\nu)}]$, is that it is possible to use conformal perturbation theory to write an expansion for the latter in terms of the deformation $\delta g_{\mu\nu} = 2\pa_{(\mu}\zeta_{\nu)}$ \cite{Rosenhaus:2014woa, Rosenhaus:2014zza, Allais:2014ata}. To see how this works, consider the reduced density matrix $\brho$ on $A$ in the presence of the metric deformation $\delta g_{\mu\nu}$. A straightforward calculation shows  (see Appendix A for details)
\beqn\label{PEDM}
\brho &=& \brho_0+\frac{1}{2}\int d^{d}\bx\;\delta g_{\mu\nu}(\bx)\;\brho_0\Big(\hT^{\mu\nu}(\bx)-\mathrm{tr}\left(\brho_0\hT^{\mu\nu}(\bx)\right)\Big)\\
&+&\frac{1}{8}\int  d^{d}\bx_a d^{d}\bx_b\;\delta g_{\mu\nu}(\bx_a)\delta g_{\lambda\sigma}(\bx_b) \;\brho_0\Big\{\mathcal{T}\left[\hT^{\mu\nu}(\bx_a)\hT^{\lambda\sigma}(\bx_b)\right]-2\hT^{\mu\nu}(\bx_a)\;\mathrm{tr}\left(\brho_0\hT^{\lambda\sigma}(\bx_b)\right)\nonumber\\
&-&\mathrm{tr}\left(\brho_0\mathcal{T} \left[\hT^{\mu\nu}(\bx_a)\hT^{\lambda\sigma}(\bx_b)\right]\right)+2\mathrm{tr}\left(\brho_0\hT^{\mu\nu}(\bx_a)\right)\;\mathrm{tr}\left(\brho_0\hT^{\lambda\sigma}(\bx_b)\right)\Big\}+\cdots\nonumber
\eeqn
where $\bx^{\mu}=(x_E^0,\bdx)$ are now coordinates on Euclidean space $\re^d$, and
\beq
\brho_0 = \frac{e^{-2\pi \hK}}{\mathrm{tr}\;e^{-2\pi \hK}}
\eeq
is the original reduced density matrix on $A$ in the absence of the metric perturbation.\footnote{From now on, by $\mathrm{tr}$ we mean $\mathrm{tr}_{\mathfrak{h}_A}$ unless otherwise specified.} Further,  $\mathcal{T}$ is the angular-ordering operator in the $(x_E^0,x^1)$ plane, i.e., if $\theta\in [0,2\pi)$ is the angular coordinate in the $(x_E^0,x^1)$ plane, then 
\beq
\mathcal{T}\left[ \widehat{\mathcal{O}}(\theta_a)\widehat{\mathcal{O}}(\theta_b)\right] = \widehat{\mathcal{O}}(\theta_a)\widehat{\mathcal{O}}(\theta_b) H (\theta_a-\theta_b)+\widehat{\mathcal{O}}(\theta_b)\widehat{\mathcal{O}}(\theta_a)H (\theta_b-\theta_a)
\eeq
where $H (\theta_a-\theta_b)$ is the Heaviside step function. The next step is to perturbatively expand the entanglement entropy $S_{EE} = -\mathrm{tr}(\brho\,\mathrm{ln}\,\brho)$ using eqaution \eqref{PEDM}. However, care must be taken in expanding the logarithm, because $\brho_0$ and $\delta\brho=\brho -\brho_0$ do not commute in general. In order to deal with this, we use the following integral representation for the entanglement entropy 
\beq
S_{EE} = -\mathrm{tr}\;\brho\; \mathrm{ln}\;\brho=\int_0^{\infty}d\beta\Big\{\mathrm{tr}\left(\frac{\brho}{\brho+\beta}\right)-\frac{1}{1+\beta}\Big\}.
\eeq
Expanding this out to second order in $\delta\brho$, we obtain
\beq\label{PEEE}
\delta S_{EE} =  \int_{0}^{\infty}d\beta\; \mathrm{tr}\left(\delta\brho\;\frac{\beta}{(\brho_0+\beta)^2}\right)-\int_0^{\infty}d\beta\;\mathrm{tr}\left(\frac{\beta}{(\brho_0+\beta)^2}\delta\brho\frac{1}{(\brho_0+\beta)}\delta\brho\right)+\cdots.
\eeq
Substituting equation \eqref{PEDM} in \eqref{PEEE}, we find that $\delta S_{EE}$ can be written as a sum of two terms
\beq\label{twoterms}
\delta S_{EE}=\delta S_{EE}^{(1)}+\delta S_{EE}^{(2)}
\eeq
coming respectively from the first and the second term in equation \eqref{PEEE}. The first term (after performing the $\beta$ integration) is given by
\beqn\label{mHterm1}
\delta S^{(1)}_{EE} &=& \frac{1}{2}\int d^d\bx\;\delta g_{\mu\nu}(\bx)\;\mathrm{tr}_{\mathrm{conn.}}\left(\brho_0\hT^{\mu\nu}(\bx)\widehat{H}_E\right)\nonumber\\
&+&\frac{1}{8}\int d^d\bx_a d^d\bx_b\;\delta g_{\mu\nu}(\bx_a)\delta g_{\lambda\sigma}(\bx_b) \mathrm{tr}_{\mathrm{conn.}}\Big(\brho_0\mathcal{T}\left[\hT^{\mu\nu}(\bx_a)\hT^{\lambda\sigma}(\bx_b)\right]\widehat{H}_E\Big)
\eeqn
where $\widehat{H}_E$ is the modular Hamiltonian corresponding to $\brho_0$ and $\mathrm{tr}_{\mathrm{conn.}}$ is the connected trace. From the above equation, we see that $\delta S_{EE}^{(1)}$ can be interpreted as the change in the expectation value of the (original) modular Hamiltonian; we will henceforth refer to this term as the modular Hamiltonian term. Given that all the operators inside the trace are naturally $\mathcal{T}$-ordered, we can rewrite the above traces in terms of connected Euclidean correlation functions
\beqn\label{mHterm2}
\delta S^{(1)}_{EE} &=& \frac{1}{2}\int d^d\bx\;\delta g_{\mu\nu}(\bx)\;\left\langle \hT^{\mu\nu}(\bx)\widehat{H}_E\right\rangle_{\mathrm{conn.}}\nonumber\\
&+&\frac{1}{8}\int d^d\bx_a d^d\bx_b\;\delta g_{\mu\nu}(\bx_a)\delta g_{\lambda\sigma}(\bx_b) \left\langle \hT^{\mu\nu}(\bx_a)\hT^{\lambda\sigma}(\bx_b)\widehat{H}_E\right\rangle_{\mathrm{conn.}}.
\eeqn

Now, the second term in \eqref{twoterms} is given by
\beqn\label{relent0}
\delta S^{(2)}_{EE}&=&-\frac{1}{4}\int d^d\bx_a d^d\bx_b\;\delta g_{\mu\nu}(\bx_a)\delta g_{\lambda\sigma}(\bx_b)\int_{0}^{\infty}d\beta\beta\\
&\times& \mathrm{tr}\,\Big\{\frac{\brho_0}{(\brho_0+\beta)^2}\left[\hT^{\mu\nu}(\bx_a)-\mathrm{tr}\left(\brho_0\hT^{\mu\nu}(\bx_a)\right)\right]\frac{\brho_0}{\brho_0+\beta}\left[\hT^{\lambda\sigma}(\bx_b)-\mathrm{tr}\left(\brho_0\hT^{\lambda\sigma}(\bx_b)\right)\right]\Big\}. \nonumber
\eeqn
This term is in fact the negative of the relative entropy of $\brho$ with respect to $\brho_0$ at second order in $\delta g_{\mu\nu}$
\beq
\delta S^{(2)}_{EE} = -S(\brho || \brho_0)=\mathrm{tr}\left(\brho\ \mathrm{ln}\;\brho_0\right) -\mathrm{tr}\left(\brho\ \mathrm{ln}\;\brho\right)
\eeq
and will henceforth be referred to as the relative entropy term. The non-negativity of relative entropy then implies
\beq
\delta S^{(2)}_{EE} \leq 0.
\eeq
Unfortunately, the operators appearing in equation \eqref{relent0} are not manifestly $\mathcal{T}$-ordered, and so the trace in this form cannot be written as a Euclidean correlation function. However, it is possible to perform the $\beta$ integral and manipulate this expression further to bring it to a more convenient form (see Appendix B)
\beqn\label{relent1}
\delta S^{(2)}_{EE} &=& \frac{1}{32}\int\;d^d\bx_a d^d\bx_b\;\delta g_{\mu\nu}(\bx_a)\delta g_{\lambda\sigma}(\bx_b)\\
&\times &\int_{-\infty}^{\infty}\frac{ds}{\mathrm{sinh}^2(s/2+i\varepsilon\mathrm{sgn}(\theta_a-\theta_b))}{\left(R^{-1}( is)\right)^{\lambda}}_{\kappa}{\left(R^{-1}( is)\right)^{\sigma}}_{\eta}\mathrm{tr}_{\mathrm{conn.}}\Big(\brho_0\mathcal{T}[\hT^{\mu\nu}(\bx_a)\hT^{\kappa\eta}(R( is)\cdot \bx_b)]\Big)\nonumber
\eeqn
where ${\left(R(\theta)\right)^{\mu}}_{\nu}$ is a rotation in the $(x^0_E, x^1)$ plane by the angle $\theta$.  This manipulation essentially involves steps similar to passing from  old-fashioned perturbation theory to time dependent perturbation theory in quantum mechanics. This is usually achieved using Schwinger parameters and the $s$ appearing above can be thought of as such. 

The trade-off however is the additional $s$ integral with the attendant measure. Interestingly, note that the way $s$ appears in the above correlation function corresponds to a relative boost between the two stress tensor insertions, with $s$ being the boost angle/rapidity. Equivalently, from the point of view of the modular Hamiltonian, we are forced into ``real time'' evolution. Indeed, we can rewrite the above equation in the following way to make this point manifest 
\beqn
\delta S^{(2)}_{EE} &=& \frac{1}{32}\int\;d^d\bx_a d^d\bx_b\;\delta g_{\mu\nu}(\bx_a)\delta g_{\lambda\sigma}(\bx_b)\\
&\times &\int_{-\infty}^{\infty}\frac{ds}{\mathrm{sinh}^2( s/2+i\varepsilon\mathrm{sgn}(\theta_a-\theta_b))}\;\mathrm{tr}_{\mathrm{conn.}}\Big(\brho_0\mathcal{T}[\hT^{\mu\nu}(\bx_a)\brho_0^{\; -is/2\pi\;}\hT^{\lambda\sigma}(\bx_b)\brho_0^{\; is/2\pi}]\Big). \nonumber
\eeqn

Having written this term in the above form we can now use the $\mathcal{T}$-ordering to rewrite the trace in terms of the Euclidean two-point correlation function to obtain
\beqn\label{relent}
\delta S^{(2)}_{EE} &=& \frac{1}{32}\int\;d^d\bx_a d^d\bx_b\;\delta g_{\mu\nu}(\bx_a)\delta g_{\lambda\sigma}(\bx_b)\\
&\times &\int_{-\infty}^{\infty}\frac{ds}{\mathrm{sinh}^2( s/2+i\varepsilon \mathrm{sgn}(\theta_a-\theta_b))}\;{\left(R^{-1}( is)\right)^{\lambda}}_{\kappa}{\left(R^{-1}(is)\right)^{\sigma}}_{\eta}\left\langle\hT^{\mu\nu}(\bx_a)\hT^{\kappa\eta}(R( is)\cdot \bx_b)\right\rangle_{\mathrm{conn.}}. \nonumber
\eeqn

From equations \eqref{mHterm2} and \eqref{relent} we see that the entanglement density can be computed in terms of the two-point and three-point Euclidean correlation functions of the stress tensor. Indeed, in any conformal field theory, these correlators are universal and fixed by conformal invariance modulo finitely many parameters \cite{Osborn:1993cr}. The two-point function in particular takes the form
\beq
\left\langle \hT_{\mu\nu}(\bx)\hT_{\lambda\sigma}(0)\right\rangle_{\mathrm{conn.}} = \frac{C_T}{|\bx|^{2d}}\left(\frac{1}{2}I_{\mu\lambda}I_{\nu\sigma}+\frac{1}{2}I_{\mu\sigma}I_{\nu\lambda}-\frac{1}{d}\delta_{\mu\nu}\delta_{\lambda\sigma}\right)
\eeq
\beq
I_{\mu\nu} = \delta_{\mu\nu}-2\frac{\bx_{\mu}\bx_{\nu}}{\bx^2}
\eeq
and is determined entirely by specifying the single parameter $C_T$. The three-point function is more complicated, and in general dimension depends on three independent parameters. Nevertheless, it is clear from the above discussion that the (non-local part of the) entanglement density in a CFT is uniquely determined in terms of the parameters appearing in the two- and three-point correlators. 

All that remains now is to explicitly evaluate the integrals in equations \eqref{mHterm2} and \eqref{relent}. Doing so, one encounters the following surprising result -- \emph{the modular Hamiltonian term \eqref{mHterm2} does not contribute to the non-local part of the entanglement density}. Since the explicit computation is somewhat tedious, we will defer the details to Section \ref{modHam}. 
We also give a quicker more sketchy proof of the vanishing of the modular Hamiltonian term, using a slightly different setup, in Appendix E.
The non-trivial contribution to the entanglement density then comes entirely from the relative entropy term. Indeed, this is why the result \eqref{planes} for the non-local part of the entanglement density depends only on the single parameter $C_T$.  We now proceed to compute the relative entropy term. 

\subsection{Relative entropy term}
\label{sec:rel}
In order to compute the integrals in \eqref{relent}, it is much more efficient to use the conformal transformation from $\mathcal{H}=S^1\times \mathbb{H}^{d-1}$ to $\re^d$ to pull-back and evaluate the integrals on $\mathcal{H}$. To see how this works, let us coordinatize $S^1\times \mathbb{H}^{d-1}$ by $y^{\alpha} = (\tau, z, x^i)$, where $\tau$ is periodic with period $2\pi$, and $(z, x^i)$ are Poincar\'e coordinates on the hyperbolic space $\mathbb{H}^{d-1}$. The metric on $\mathcal{H}$ in these coordinates is given by
\beq
g_{\mathcal{H}} = d\tau^2+\frac{dz^2+\delta_{ij}dx^idx^j}{z^2}.
\eeq
The map $\varphi:\mathcal{H} \to \re^d$ given by
\beq
\varphi(\tau,z,x^i) = \left(z\sin\tau, z\cos\tau, x^i\right)
\eeq
is a conformal transformation, i.e.
\beq
\varphi_*g_{\re^d}=\Omega^2(y)\;g_{\mathcal{H}}
\eeq 
with $\Omega(y)=z$ being the Weyl factor (and $\varphi_*$ being the pullback). This implies that the stress tensors on the two spaces are related by
\beq
\hT^{\mu\nu}(\bx)=\frac{\pa \bx^{\mu}}{\pa y^{\alpha}}\frac{\pa \bx^{\nu}}{\pa y^{\beta}}\hT^{\alpha\beta}(y)\Omega^{-2-d}(y)+\mathcal{S}^{\mu\nu}
\eeq
where $\mathcal{S}^{\mu\nu}$ denotes additional Schwarzian derivative-type terms, which vanish in odd dimensions, but are present in even dimensions.
The integral \eqref{relent} then pulls back to
\begin{align}\label{relent2}
\delta S^{(2)}_{EE} & = \frac{1}{32}\int\;d\mu_a d\mu_b\;h_{\alpha\beta}(y_a)\Omega^{-2}  (y_a)h_{\gamma\delta}(y_b)\Omega^{-2}(y_b) \Pi^{\alpha\beta\gamma\delta}(y_a,y_b)  \\ &  {\rm where} \qquad    \Pi^{\alpha\beta\gamma\delta} (y_a,y_b)  =  \int_{-\infty}^{\infty}\frac{ds}{\mathrm{sinh}^2( s/2+i\varepsilon\ \mathrm{sgn}(\tau_a-\tau_b))}\;\left\langle\hT^{\alpha\beta}(y_a)\hT^{\gamma\delta}(y_b^s)\right\rangle_{\mathcal{H}}
\label{using}
\end{align}
where we have defined $y_b^s = (\tau_b+ is,z_b, x_b^i)$, and $d\mu = \sqrt{\det g_{\mathcal{H}}(y)}d^dy$ is the measure on $\mathcal{H}$. Further
\beq\label{dg}
h_{\alpha\beta}\Omega^{-2} = 2\nabla_{(\alpha}\xi_{\beta)}+g^{\mathcal{H}}_{\alpha\beta}\xi^{\gamma}\pa_{\gamma}\mathrm{ln}(\Omega^2)
\eeq
where the vector field $\xi^{\alpha}$ on $\mathcal{H}$ is the push-forward of the vector field $\zeta^{\mu}$ on $\re^{d}$ by $\varphi^{-1}$
\beq
\xi = \left(\zeta^1(x^i)\cos(\tau)+\zeta^{0}(x^i)\sin(\tau)\right)\pa_{z}+\frac{1}{z}\left(-\zeta^1(x^i)\sin(\tau)+\zeta^{0}(x^i)\cos(\tau)\right)\pa_{\tau}.
\eeq
Note that the Schwarzian terms $\mathcal{S}^{\mu\nu}$ have dropped out of equation \eqref{relent2} because of the connectedness of the correlation function. Additionally, we note that the second term in \eqref{dg} can also be dropped by the tracelessness of the stress tensor (more precisely, the trace Ward identity) in conformal field theories.\footnote{In even dimensions, there are contributions coming from the trace anomaly. However, these contributions are local at the present order. Since we are interested in the non-local part of the entanglement density, we can drop these terms.} So we obtain
\beq\label{relent2.5}
\delta S^{(2)}_{EE} = \frac{1}{8}\int_{\mathcal{H}}d\mu_a\int_{\mathcal{H}} d\mu_b\;\nabla_{(\alpha}\xi_{\beta)}(y_a)\nabla_{(\gamma}\xi_{\delta)}(y_b)\Pi^{\alpha\beta\gamma\delta} (y_a,y_b).
\eeq
Since the above integrals include integration over hyperbolic space, there are potential divergences coming from the conformal boundary of hyperbolic space at $z = 0$. These divergences in the entanglement entropy of course correspond to the short-range entanglement coming from the region close to the entangling surface. One way to regulate such potential divergences is to put a cut-off at $z= \frac{1}{\Lambda}$ (which corresponds to cutting out a tubular neighbourhood around the entangling surface in the original description on Euclidean space). We denote the resulting regulated space as $\mathcal{H}_{\Lambda}$, and rewrite the above integral as
\beq\label{relent3}
\delta S^{(2)}_{EE} = \frac{1}{8}\int_{\mathcal{H}_{\Lambda}}d\mu_a\int_{\mathcal{H}_{\Lambda}} d\mu_b\;\nabla_{(\alpha}\xi_{\beta)}(y_a)\nabla_{(\gamma}\xi_{\delta)}(y_b)
\Pi^{\alpha\beta\gamma\delta} (y_a,y_b).
\eeq
Next, integrating by parts and using the diffeomorphism Ward identity,\footnote{
Which says that $\nabla_{\alpha}^a \Pi^{\alpha\beta\gamma\delta}(y_a,y_b) = 0$ and $\nabla_{\gamma}^b \Pi^{\alpha\beta\gamma\delta}(y_a,y_b) = 0$ for separated points.} 
we arrive at 
\beq\label{relent4}
\delta S^{(2)}_{EE} = \frac{1}{8}\int_{\pa\mathcal{H}_{\Lambda}}d\bar{\mu}_a n^{\alpha}(y_a)\xi^{\beta}(y_a)\int_{\pa\mathcal{H}_{\Lambda}} d\bar{\mu_b} n^{\gamma}(y_b)\;\xi^{\delta}(y_b)
\Pi_{\alpha\beta\gamma\delta} (y_a,y_b)
\eeq
where $\pa\mathcal{H}_{\Lambda}$ is the boundary of the regulated space $\mathcal{H}_{\Lambda}$ at $z=\frac{1}{\Lambda}$, $n=\frac{1}{\Lambda}\pa_z$ is the outward pointing unit-normal on the boundary, and $d\bar{\mu}$ is the measure induced on the boundary
\beq
d\bar{\mu}\; n^{\alpha}(y)= d\tau d^{d-2}x^i\;\sqrt{\det\gamma_{\pa\mathcal{H}_{\Lambda}}}\;n^{\alpha}(y)=d\tau d^{d-2}x^i\;\Lambda^{(d-2)}\;\frac{\delta^{\alpha}_{z}}{\Lambda}.
\eeq
The next thing to compute is the two-point function of stress tensors on $\mathcal{H}$. This can be done efficiently using the embedding space formalism developed in \cite{Costa:2011mg}. In this formalism (see Section~2 of \cite{Faulkner:2014jva} for a review relevant for this calculation), one considers the larger embedding space (or ambient space) $\re^{1,d+1}$ on which the (Euclidean) conformal group acts linearly. Let us pick global coordinates $P^A = (P^I,P^{II}, P^0,\cdots, P^{d-1})$ on this space, with the coordinate $P^I$ being time-like. One then embeds $\mathcal{H}$ (more generally, any space which is conformally equivalent to $\re^d$) as a section of the upper light-cone $P^2=0,\;P^I>0$. Here, we pick the embedding
\beq\label{section}
P^A(y) = \left(\frac{1+z^2+\delta_{ij}x^ix^j}{2z}, \cos(\tau),\sin(\tau), \frac{1-z^2-\delta_{ij}x^ix^j}{2z},\frac{x^i}{z}\right).
\eeq
Now the two-point function of stress tensors in equation \eqref{using} can be computed using the embedding space formalism following \cite{Costa:2011mg},
\beq
\left\langle\hT_{\alpha\beta}(y_a)\hT_{\gamma\delta}(y_b^s)\right\rangle_{\mathcal{H}_\Lambda}=\frac{1}{4}\mathbb{P}_{\alpha\beta}^{AB}(P_a)\mathbb{P}_{\gamma\delta}^{CD}(P_b^s)\frac{\pa}{\pa Z_a^A}\frac{\pa}{\pa Z_a^B}\frac{\pa}{\pa Z_b^C}\frac{\pa}{\pa Z_b^D}G^{(2)}(P_a,P^s_b,Z_a,Z_b)
\eeq
where $Z_{a,b}$ are auxiliary variables, and the right hand side above is evaluated on the section \eqref{section}. The index-free function $G^{(2)}$ is defined as,
\beq
G^{(2)}(P_a,P^s_b,Z_a,Z_b)= \frac{4C_T}{(-2P_a\cdot P_b^s)^{d+2}}\Big((P_a\cdot P_b^s)(Z_a\cdot Z_b)-(P_a\cdot Z_b)(P_b^s \cdot Z_a)\Big)^2.
\eeq
Further, the projector $\mathbb{P}^{AB}_{\alpha\beta}$ is defined as (see equations \eqref{proj1} and \eqref{proj2} for explicit expressions)
\beq
\mathbb{P}^{AB}_{\alpha\beta}(P) = \frac{\pa P^{(A}}{\pa y^{\alpha}}\frac{\pa P^{B)}}{\pa y^{\beta}}-\frac{1}{d} \eta^{AB}\eta_{CD}\frac{\pa P^{C}}{\pa y^{\alpha}}\frac{\pa P^{D}}{\pa y^{\beta}}.
\eeq
Using this formalism, we compute the required two-point correlation functions
\beq\label{cor1}
\left\langle\hT_{z\tau}(y_a)\hT_{z\tau}(y_b^s)\right\rangle_{\mathcal{H}_\Lambda}= \frac{C_T|x_a-x_b|^4\Big(\cos(2(\tau_a-\tau_b- is))+O\left(\Lambda^{-2}\right)\Big)}{2\Lambda^{2d-2}\Big(|x_a-x_b|^2+\frac{2}{\Lambda^2}\left(1-\cos(\tau_a-\tau_b- is)\right)\Big)^{d+2}}
\eeq
\beq\label{cor2}
\left\langle\hT_{z\tau}(y_a)\hT_{zz}(y_b^s)\right\rangle_{\mathcal{H}_\Lambda}= -\frac{C_T|x_a-x_b|^4\Big(\sin(2(\tau_a-\tau_b-is))+O\left(\Lambda^{-2}\right)\Big)}{2\Lambda^{2d-3}\Big(|x_a-x_b|^2+\frac{2}{\Lambda^2}\left(1-\cos(\tau_a-\tau_b- is)\right)\Big)^{d+2}}
\eeq
\beq\label{cor3}
\left\langle\hT_{zz}(y_a)\hT_{zz}(y_b^s)\right\rangle_{\mathcal{H}_\Lambda}= \frac{C_T|x_a-x_b|^4\Big(\cos(2(\tau_a-\tau_b- is))+\frac{d-2}{d}+O\left(\Lambda^{-2}\right)\Big)}{2\Lambda^{2d-4}\Big(|x_a-x_b|^2+\frac{2}{\Lambda^2}\left(1-\cos(\tau_a-\tau_b- is)\right)\Big)^{d+2}}
\eeq
where $|x_a-x_b|^2=\delta_{ij}(x_a-x_b)^i(x_a-x_b)^j$. The $O(\Lambda^{-2})$ terms in the above expressions do not contribute in the limit $\Lambda \to \infty$, so we will drop them henceforth. Substituting equations \eqref{cor1}--\eqref{cor3} in \eqref{relent4} and using \eqref{using}, we obtain
\beq\label{relent5}
\delta S^{(2)}_{EE}=\frac{ C_T}{16\Lambda^2}\int_0^{2\pi} d\tau_a \int d^{d-2}x_a\int_0^{2\pi}d\tau_b\int d^{d-2}x_b\;|x_a-x_b|^4\hat{\xi}_a^{\ui}\mathcal{M}_{\ui \uj}\hat{\xi}_b^{\uj}
\eeq
where the overlined indices run over $\ui, \uj = (\tau, z)$, and
\beq
\hat{\xi}= \left(\zeta^1\cos(\tau)+\zeta^{0}\sin(\tau)\right)\pa_{z}+\left(-\zeta^1\sin(\tau)+\zeta^{0}\cos(\tau)\right)\pa_{\tau}.
\eeq
We have also defined 
\beq\label{M1}
\mathcal{M}_{\ui\uj}=\int_{-\infty}^{\infty}\frac{ds}{\mathrm{sinh}^2(s/2+i\varepsilon \mathrm{sgn}(\tau_a-\tau_b))}\frac{1}{\Big(|x_a-x_b|^2+\frac{2}{\Lambda^2}\left(1-\cos(\tau_a-\tau_b- is)\right)\Big)^{d+2}}M_{\ui\uj}
\eeq
with $M_{\ui\uj}$ given by the two dimensional matrix
\beq\label{M2}
M_{\ui\uj} = \left(\begin{matrix}\cos\left(2(\tau_a-\tau_b- is)\right) & -\sin\left(2(\tau_a-\tau_b- is)\right)\\ & \\\sin\left(2(\tau_a-\tau_b- is)\right) & \cos\left(2(\tau_a-\tau_b- is)\right)+\frac{d-2}{d}\end{matrix}\right).
\eeq
The factor of $\frac{1}{\Lambda^2}$ out front in equation \eqref{relent5} apparently suppresses $\delta S_{EE}^{(2)}$ in the limit $\Lambda \to \infty$. However, we must be careful in taking this limit because there is a possible enhancement from the $s$ integration inside $\mathcal{M}_{\ui\uj}$. Indeed, naively sending $\Lambda \to \infty$ inside the integral in equation \eqref{M1}, we see that the $s$ integral diverges as $\int ds\ e^{s}$ in the limits $s\to \pm \infty$. We can extract these divergences by zooming in on the integral in these limits; this gives two contributions
\beq
\mathcal{M}_{\ui\uj}=\mathcal{M}^{\infty}_{\ui\uj}+\mathcal{M}^{-\infty}_{\ui\uj}.
\eeq
The contribution from $s\to \pm\infty$ can be extracted by changing variables to $\beta= \Lambda^{-2} e^{\pm s}$
\beq
\mathcal{M}^{\pm\infty}_{\ui\uj} \simeq 2\Lambda^2 \int_0^{\infty}d\beta\frac{e^{\pm 2i(\tau_a-\tau_b)}}{\Big(|x_a-x_b|^2-\beta e^{\pm i(\tau_a-\tau_b)}\Big)^{d+2}}\left(\begin{matrix}1 & \pm i\\ \mp i & 1\end{matrix}\right)=-\frac{2\Lambda^2}{ (d+1)}\frac{e^{\pm i(\tau_{a}-\tau_b)}}{|x_a-x_b|^{2(d+1)}}\left(\begin{matrix}1 & \pm i\\ \mp i & 1\end{matrix}\right)
\eeq
Finally substituting the above into equation \eqref{relent5} and integrating over $\tau_a,\tau_b$, we get
\beq
\delta S_{EE}^{(2)} = -\frac{2\pi^2C_T}{(d+1)}\int d^{d-2}x_ad^{d-2}x_b\frac{1}{2}\left(\zeta^1(x_a)\zeta^1(x_b)+\zeta^0(x_a)\zeta^0(x_b)\right)\frac{1}{|x_a-x_b|^{2(d-1)}}.
\eeq
Reverting back to Lorentzian signature, we obtain 
\beq
\delta S_{EE}^{(2)} = -\frac{2\pi^2C_T}{(d+1)}\int d^{d-2}x_ad^{d-2}x_b\;\frac{1}{2}\zeta^{\ui}(x_a)\frac{\eta_{\ui\uj}}{|x_a-x_b|^{2(d-1)}}\zeta^{\uj}(x_b)
\eeq
which is our primary result. It still remains to be shown however, that the modular Hamiltonian term $\delta S_{EE}^{(1)}$ does not give additional contributions -- we show this in the next section. 

To end this section we would like to give some insight into the above calculation and in particular where
the main contribution to the non-local part of the entanglement density is coming from. In words, the two
stress tensor insertions start their lives close to the boundary of the entangling region
(a distance $1/\Lambda$ from $\partial A$ in the flat space metric.) However when we boost one of these operators by  a rapidity of order $ 2 \ln \Lambda$,  then the stress tensor gets liberated from $\partial A$ and moves into one of the null generators of $\partial \mathcal{D}(A)$, the boundary of the domain of dependence of $A$ --- otherwise known as the Rindler horizon. Here the relevant integrated correlation function receives an enhancement of order $\Lambda^2$ in such a way that $\Lambda$ drops out of the final expression. Thus the main contribution to the entanglement density is coming from the correlation function of stress tensors inserted along null generators of $\partial \mathcal{D}(A)$.
We find this result intriguing and intend to study this further in future works. 

\subsection{The modular Hamiltonian term} \label{modHam}
Now we return to the modular Hamiltonian term -- in particular, the second term in equation \eqref{mHterm2} (the first term in \eqref{mHterm2} was studied in ref. \cite{Allais:2014ata} where it was shown to have no universal contributions). An alternative, less constructive, proof that
this term vanishes is given in Appendix \ref{sec:quick}. 
 Using the map $\varphi:\mathcal{H} \to \re^d$, this term pulls back to
\beq
\delta S_{EE}^{(1)}=\frac{1}{8}\int_{\mathcal{H}_{\Lambda}} d\mu_a \int_{\mathcal{H}_{\Lambda}} d\mu_b\;\Omega^{-2}(y_a)h_{\alpha\beta}(y_a)\Omega^{-2}(y_b)h_{\gamma\delta}(y_b) \left\langle \hT^{\alpha\beta}(y_a)\hT^{\gamma\delta}(y_b)\widehat{H}_E\right\rangle_{\mathcal{H}_\Lambda}
\eeq
where recall that 
\beq
\Omega^{-2}h_{\alpha\beta}=2\nabla_{(\alpha}\xi_{\beta)}+\xi\cdot\partial\;\mathrm{ln}(\Omega^2)\;g^{\mathcal{H}}_{\alpha\beta}.
\eeq
Further, integrating by parts and using the diffeomorphism and trace Ward identities allows us to rewrite this as
\beq\label{mH0}
\delta S_{EE}^{(1)}=\frac{1}{8}\int_{\pa\mathcal{H}_{\Lambda}} d\bar{\mu}_a n^{\alpha}(y_a)\xi^{\beta}(y_a) \int_{\pa\mathcal{H}_{\Lambda}} d\bar{\mu}_b n^{\gamma}(y_b)\xi^{\delta}(y_b)\;\left\langle \hT_{\alpha\beta}(y_a)\hT_{\gamma\delta}(y_b)\widehat{H}_E\right\rangle_{\mathcal{H}_\Lambda}+\mathrm{contact\;terms}.
\eeq
In Appendix C, we will argue that the contact terms vanish in the limit $\Lambda \to \infty$, and so we focus here on the three-point function term.  The modular Hamiltonian $\widehat{H}_E$ written on $S^1\times \mathbb{H}^{d-1}$ is simply the generator of $\tau$-translations
\beq\label{mH}
\widehat{H}_E=2\pi\int_{\mathbb{H}^{d-1}}\frac{d^{d-2}x^i_cdz_c}{z_c^{d-1}}\;\hT_{\tau\tau}(\tau_c,z_c,x^i_c)+\mathrm{constant}
\eeq
where the integral above is on the constant $\tau=\tau_c$ slice. The constant term above drops out of all connected correlators. So the relevant correlation function in the present calculation is the three-point function of stress tensors on $\mathcal{H}$
\beq\label{3pt1}
\delta S_{EE}^{(1)}=\frac{\pi}{4}\int_{\pa\mathcal{H}_{\Lambda}} d\bar{\mu}_a n^{\alpha}(y_a)\xi^{\beta}(y_a) \int_{\pa\mathcal{H}_{\Lambda}} d\bar{\mu}_b n^{\gamma}(y_b)\xi^{\delta}(y_b)\int_{\mathbb{H}^{d-1}}\frac{d^{d-2}x^i_cdz_c}{z_c^{d-1}}\;\left\langle \hT_{\alpha\beta}(y_a)\hT_{\gamma\delta}(y_b)\hT_{\tau\tau}(y_c)\right\rangle_{\mathcal{H}_\Lambda}.
\eeq

Once again, it is efficient to use the embedding space formalism to obtain this correlation function
\beq\label{3ptES}
\left\langle \hT_{\alpha\beta}(y_a)\hT_{\gamma\delta}(y_b)\hT_{\tau\tau}(y_c)\right\rangle_{\mathcal{H}_\Lambda}=\frac{1}{8}\mathbb{P}^{AB}_{\alpha\beta}(P_a)\mathbb{P}^{CD}_{\gamma\delta}(P_b)\mathbb{P}^{EF}_{\tau\tau}(P_c)\frac{\pa^2}{\pa Z^A_a\pa Z^B_a}\frac{\pa^2}{\pa Z^C_b\pa Z^D_b}\frac{\pa^2}{\pa Z^E_c\pa Z^F_c} G^{(3)}
\eeq
where\footnote{Additionally, there are further contact terms in the three-point function which are required by the trace Ward identity. However, these contact terms do not contribute in the limit $\Lambda \to \infty$, and so we do not show them here explicitly.} 
\beq\label{g3}
G^{(3)}(P_a,P_b,P_c;Z_a,Z_b, Z_c)=(-2P_a\cdot P_b)^{-\frac{d+2}{2}}(-2P_b\cdot P_c)^{-\frac{d+2}{2}}(-2P_c\cdot P_a)^{-\frac{d+2}{2}}\sum_{m=1}^{5}\alpha_mA_m(V_I,H_{JK}).
\eeq
The $\{A_m\}$'s are conformally invariant structures -- polynomials made from six basic building blocks \cite{Costa:2011mg}
\beq
H_{IJ}=-2\Big((Z_I\cdot Z_J)(P_I\cdot P_J)-(Z_I\cdot P_J)(Z_J\cdot P_I)\Big)
\eeq 
\beq
V_I= \frac{(Z_I\cdot P_J)(P_I\cdot P_K)-(Z_I\cdot P_K)(P_I\cdot P_J)}{P_J\cdot P_K}
\eeq
where the triplet $(I,J,K)$ is a cyclic permutation of $(a,b,c)$. The allowed structures are
\footnote{In three dimensions there is also potentially a parity odd structure \cite{Maldacena:2011nz,Giombi:2011rz} that we did not write down. However it is easy to argue that such a term cannot contribute
to the non-local part of the entanglement density based on symmetries and unitarity - there is no parity
odd term that we could add to \eqref{planes} or \eqref{spheres} which preserves the appropriate conformal symmetries and the strong subaddativity constraint. }
\beq
A_1= V_a^2V_b^2V_c^2
\eeq
\beq
A_2= H_{ab}V_aV_bV_c^2+H_{bc}V_bV_cV_a^2+H_{ca}V_cV_aV_b^2
\eeq
\beq
A_3= V_aH_{ab}H_{bc}V_c+V_bH_{bc}H_{ca}V_a+V_cH_{ca}H_{ab}V_b
\eeq
\beq
A_4 = H_{ab}^2V_c^2+H_{bc}^2V_a^2+H_{ca}^2V_b^2
\eeq
\beq
A_5= H_{ab}H_{bc}H_{ca}.
\eeq
The coefficients $\alpha_m$ in (\ref{g3}) are not all independent -- imposing the conservation condition on the stress tensors for non-coincident points gives the constraints
\beq\label{cc1}
C_1(\alpha_m)\equiv -2\alpha_1+4\alpha_2+\left(\frac{d^2}{2}+d-4\right)\alpha_3-d(d+2)\alpha_4=0
\eeq
\beq\label{c2}
C_2(\alpha_m)\equiv \alpha_2-\frac{d+2}{2}\alpha_3+2d\alpha_4+\frac{1}{2}(4-d^2)\alpha_5=0.
\eeq
This fixes two of the coefficients in terms of the rest, leaving three independent coefficients.\footnote{Imposing the conservation equation in the coincident limit fixes a linear combination of these three coefficients in terms of $C_T$ \cite{Osborn:1993cr}. }

In general, computing the integrals in \eqref{3pt1} over the hyperbolic slice at $\tau=\tau_c$ is a difficult task. However, the following observation makes this computation tractable -- the modular Hamiltonian is a conserved charge and so we are free to move it in $\tau$. One therefore expects $\delta S_{EE}^{(1)}$ to be independent of $\tau_c$. One might worry about potential crossing contributions to $\delta S_{EE}^{(1)}$ when we move the modular Hamiltonian across one of the other stress-tensor insertions, but it can be checked explicitly that these vanish in the limit $\Lambda \to \infty$ (see Appendix C). Therefore, in the complex $w = e^{i\tau_c}$ plane, $\delta S_{EE}^{(1)}$ can be extended to an analytic function which is constant along the unit circle $|w|=1$, and hence a constant on the entire $w$ plane. We can therefore use this to our advantage by computing $\delta S_{EE}^{(1)}$ at a special point such as $w=0$ or $w=\infty$. Physically, these two-points correspond to light-cone limits: $w\to 0$ corresponds to writing the modular Hamiltonian as an integral over the past null boundary of the domain of dependence $\mathcal{D}(A)$ (or the past Rindler horizon for brevity), while $w \to \infty$ corresponds to writing it as an integral over the future Rindler horizon (see Fig.~\ref{fig:fig3}). We take $w \to \infty$ in what follows.
\myfig{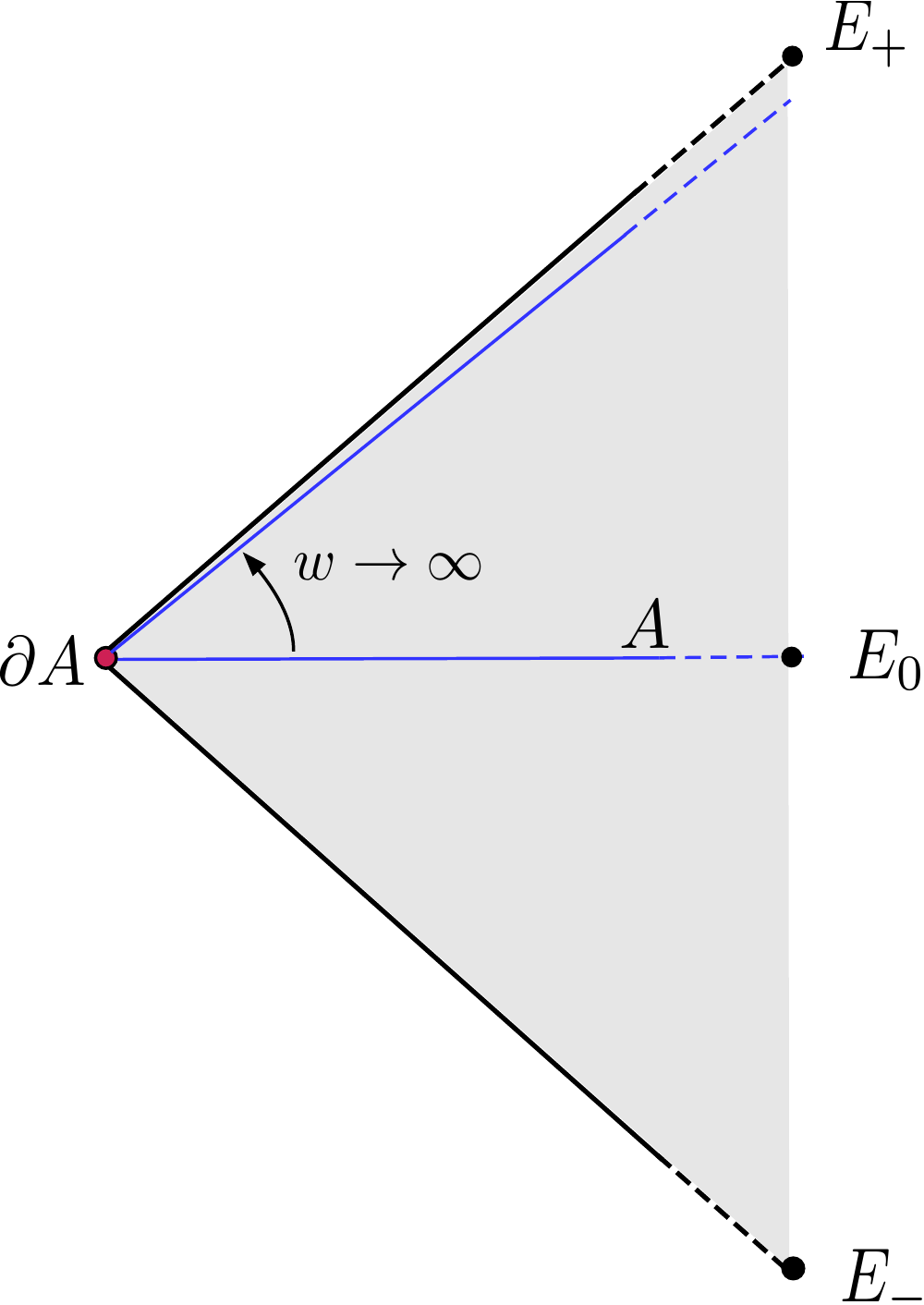}{5.7}{\textsf{The shaded region is the domain of dependence $\mathcal{D}(A)$. Analytically continuing in $w=e^{i\tau_c}$ and sending $w\to \infty$ sends the modular Hamiltonian (integrated over the blue line) to the future null boundary of $\mathcal{D}(A)$. Also shown are the future and past tips $E_{\pm}$ of $\mathcal{D}(A)$ and the point at spacelike infinity $E_0$.}}

The computation simplifies dramatically in this limit. To see this is more detail, define the points
\beq
E_{\pm} = \left(0, 1, \pm i, 0,\cdots, 0\right),\;\; E_0=(1,0,0,-1,0\cdots,0)
\eeq
in the embedding space. These points have an interesting physical interpretation --- (the rays corresponding to) $E_{\pm}$ form the future and past tips of the light cone comprising the boundary of the domain of dependence of $A$, while (the ray corresponding to) $E_0$ constitutes the point at spacelike infinity, or equivalently the point at infinity in the Poincar\'e coordinates of hyperbolic space $\mathbb{H}^{d-1}$. We note the relations
\beq
E_+\cdot E_+ = E_-\cdot E_- = 0,\;\;E_+\cdot E_- = 2,\;\;E_0^2= 0,\;\; E_0\cdot E_{\pm}=0.
\eeq
The relevant projectors in \eqref{3ptES} (with $z_{a,b}=\frac{1}{\Lambda}$) can be written in terms of these points as\footnote{The function $G^{(3)}$ in \eqref{g3}, by construction, satisfies the transversality conditions $P_{a,b,c}\cdot \frac{\pa}{\pa Z_{a,b,c}}G^{(3)}=0$, and we have used these to simplify the projectors.}
\beq\label{proj1}
\mathbb{P}^{AB}_{\tau z}(P_{a,b}) = \frac{i\Lambda}{4}\Big((P_{a,b}^+)^2E_-^AE_-^B-(P_{a,b}^-)^2E_+^AE_+^B\Big)+\frac{1}{2i}\left(P_{a,b}^-E_+^{(A}-P_{a,b}^+E_-^{(A}\right)E^{B)}_0
\eeq
\beqn\label{proj2}
\mathbb{P}^{AB}_{z z}(P_{a,b}) &=& \frac{\Lambda^2}{4}\Big((P_{a,b}^+)^2E_-^AE_-^B+(P_{a,b}^-)^2E_+^AE_+^B+2E_+^{(A}E_-^{B)}-\frac{4}{d}\eta^{AB}\Big)\\
&+&\Lambda\left(P_{a,b}^-E_+^{(A}+P_{a,b}^+E_-^{(A}\right)E_0^{B)}+E_0^AE_0^B\nonumber
\eeqn
\beq
\lim_{w \to \infty} \mathbb{P}_{\tau\tau}^{AB}(P_c) = -\frac{w^2}{4}\left(E_{\mp}^AE_{\mp}^B+O(w^{-1})\right)
\eeq
where $P^{\pm}_{a,b} = P_{a,b}\cdot E^{\pm}$. The only other ingredient required to compute $\delta S_{EE}^{(1)}$ is the following generic integral 
\beq
I(n_+,n_-,n_0|m_a,m_b)=\int_{\mathbb{H}^{d-1}} dY_c\;(-2P_a\cdot P_c)^{-m_a}(-2P_b\cdot P_c)^{-m_b}(E_+\cdot P_c)^{n_+}(E_-\cdot P_c)^{n_-}(E_0\cdot P_c)^{n_0}
\eeq
where $dY_c$ is the appropriate integration measure over $\mathbb{H}^{d-1}$. Precisely in the limit $w \to \infty$, this integral simplifies greatly and can be written in terms of a single integral (see Appendix D)
\beq
\label{lamint}
I(n_+,n_-,n_0|m_a,m_b)=-\frac{k}{w^2}\delta_{n_-,0}\int_0^{\infty}  d\lambda\; \frac{\lambda^{m_a-m_b-1}}{\left(\lambda P_a^{\mp}+\lambda^{-1}P_b^{\mp}\right)^{n_++2}}\frac{\left(-\lambda (E_0\cdot Y_a)-\lambda^{-1}(E_0\cdot Y_b)\right)^{n_0}}{(\lambda^2+\frac{1}{\lambda^2}-2Y_a\cdot Y_b)^{\frac{d}{2}-1+n_0}}+O(w^{-3})
\eeq
where $Y_{a,b}$ are the embedding space coordinates for $\mathbb{H}^{d-1}$, and the constant $k$ is given by
$$k = 2\pi^{\frac{d-2}{2}}(-1)^{n_0+n_+}\frac{\Gamma(\frac{d}{2}-1+n_0)\Gamma(n_++2)}{\Gamma(m_a)\Gamma(m_b)}.$$
Putting everything together, one finds
\beq
\delta S_{EE}^{(1)}=\frac{\pi}{4}\int_{\pa\mathcal{H}_{\Lambda}} d\bar{\mu}_a  \int_{\pa\mathcal{H}_{\Lambda}} d\bar{\mu}_b\; \hat{\xi}^{\ui}(y_a)\mathcal{N}_{\ui\uj}\hat{\xi}^{\uj}(y_b)
\eeq
where the matrix $\mathcal{N}_{\ui\uj}$ can be explicitly computed as a series expansion in $\Lambda^2$. For instance, 
\beq
\label{expl}
\mathcal{N}_{\tau\tau} = \frac{c}{|x_a-x_b|^{2(d-1)}}\Big(\Lambda^2|x_a-x_b|^2C_1(\alpha_m)-d\left[2-\cos(\tau_a-\tau_b)\right]C_1(\alpha_m)+O\left(\Lambda^{-2}\right)\Big)
\eeq
where 
\beq
c=\frac{(-1)^d\pi^{d/2-1}}{d^2(d+2)^2(d+4)\Gamma\left(\frac{d}{2}\right)}.
\eeq
Now comes the surprising part: the terms which could potentially survive in the $\Lambda \to \infty$ limit come multiplied by the function $C_1(\alpha_m)$ defined in \eqref{cc1}, which vanishes by the conservation constraints. The same is true for all the components of the matrix $\mathcal{N}$ -- the potentially non-trivial terms in the $\Lambda \to \infty$ limit are all proportional to linear combinations of $C_{1}(\alpha_m)$ and $C_2(\alpha_m) $. Therefore,
\beq
\lim_{\Lambda \to \infty}\mathcal{N}_{\ui\uj}=0.
\eeq
This completes our proof that the modular Hamiltonian term does not give additional contributions to the non-local part of the entanglement density.
\subsection{Spherical case}
\newcommand{\bX}{\boldsymbol{X}}
So far, we have presented the calculation for the non-local part of the entanglement density in the case of planar entangling surfaces. It is possible to repeat the above calculation for spherical entangling surfaces, but here we will obtain the corresponding result more directly by making use of the conformal transformation $\psi:\re^{1,d-1} \to \re^{1,d-1}$ given by
\beq
\psi^{\mu}(\bX)= \frac{\bX^{\mu}-(\bX\cdot \bX)C^{\mu}}{1-2(C\cdot \bX)+(\bX\cdot \bX)(C\cdot C)}+2R^2C^{\mu}
\eeq
where we have used $\bX^{\mu} = (X^0, X^1, X^i)$ as coordinates on the domain of $\psi$, and $C=\left(0,\frac{1}{2R},0,\cdots, 0\right)$. This is a conformal transformation because $\psi_*\eta = \omega^2\eta$, with the conformal factor $\omega$ given by
\beq
\omega(\bX) = \frac{1}{1-2(C\cdot \bX)+(\bX\cdot \bX)(C\cdot C)}.
\eeq
If we use global coordinates $\bx^{\mu} = (x^0, \bdx)$ to cover the image of this map, then it is a simple matter to check the following statements: (i) $\psi$ maps the Cauchy surface $X^0=0$ to the Cauchy surface $x^0=0$, (ii) if $A$ is the half-space $X^1\geq 0$ on the Cauchy surface $X^0=0$, then $B=\psi(A)$ is the ball-shaped region $\bdx^2 \leq R^2$ on the Cauchy surface $x^0=0$, (iii) $\psi$ maps the domain of dependence of $A$ to the domain of dependence of $B$. Consequently, we can compute the entanglement density for the ball-shaped region $B$ by pushing forward the deformation vector field $\zeta_{\pa B}(\Omega^i)$ (where $\Omega^i$ are coordinates on the sphere $\pa B$) by $\psi^{-1}$
\beq\label{pb}
\zeta_{\pa A}^{\mu}(X^i) = \frac{\pa \bX^{\mu}}{\pa \bx^{\nu}}\zeta_{\pa B}^{\nu}(\Omega^i)
\eeq
and then computing the corresponding entanglement density for the half-space $A$ 
\beq
\delta^{(2)} S_{EE}(B) =\delta^{(2)} S_{EE}(A) =  -\frac{2\pi^2C_T}{d+1}\int d^{d-2}X_ad^{d-2}X_b\;\frac{1}{2}\zeta_{\pa A}^{\ui}(X_a)\zeta_{\pa A}^{\uj}(X_b)\frac{\eta_{\ui\uj}}{|X_a-X_b|^{2(d-1)}}
\eeq
Since the map $\psi$ is a conformal transformation, the Jacobian factor in \eqref{pb} can be written as 
\beq
\frac{\pa \bX^{\mu}}{\pa \bx^{\nu}}= \omega^{-1}(\bX)\;{\mathcal{R}^{\mu}}_{\nu}(\psi)
\eeq
where $\mathcal{R}$ is a rotation. Since $\zeta_{\pa B}$ lies in the plane perpendicular to the entangling surface $\bdx^2=R^2$, it follows that $\zeta_{\pa A}$ also lies in the plane perpendicular to the surface $X^1=0$. Further, by rotation symmetry along $\pa B$ (or equivalently, translation symmetry along its inverse image $\pa A$), we deduce
\beq
\eta_{\ui\uj}\;\zeta_{\pa A}^{\ui}(X_a)\zeta_{\pa A}^{\uj}(X_b)=\omega^{-1}(0,0,X_a)\omega^{-1}(0,0,X_b)\eta_{\ui\uj}\;\zeta_{\pa B}^{\ui}(\Omega_a)\zeta_{\pa B}^{\uj}(\Omega_b)
\eeq
Finally, using the relations
\beq
R^{d-2}d^{d-2}\Omega_a = \omega^{d-2}(0,0,X_a)d^{d-2}X_a
\eeq
\beq
(\bx_a-\bx_b)^2 = \omega(\bX_a)\omega(\bX_b)(\bX_a-\bX_b)^2
\eeq
we obtain 
\beq
\delta^{(2)} S_{EE}(B) = -\frac{2\pi^2C_T}{R^2(d+1)}\int d^{d-2}\Omega_a d^{d-2}\Omega_b\;\frac{1}{2}\zeta_{\pa B}^{\ui}(\Omega_a)\zeta_{\pa B}^{\uj}(\Omega_b)\frac{\eta_{\ui\uj}}{|\Omega_a-\Omega_b|^{2(d-1)}}
\eeq
which is the result for the entanglement density for spheres.
\section{Applications}\label{app}
In this section we will present some applications of our formula for the entanglement density. In section \ref{3dcorners}, we will prove the conjectured universality of the corner term contribution to entanglement entropy in $d=3$ \cite{Bueno:2015rda, Bueno:2015xda}. In section \ref{Mezei}, we will prove the Mezei formula for the shape dependence of entanglement entropy across deformed spheres, which was conjectured based on holographic calculations in \cite{Mezei:2014zla}. In \cite{Bueno:2015lza}, the Mezei formula was used to deduce further universality results for corner terms in higher dimensions. Our proof of the Mezei formula thus also establishes the universality of corner terms in higher dimensions.

\subsection{Corner terms in $d=3$}\label{3dcorners}
\label{sec:corner}
The entanglement entropy of a general subregion in the vacuum state of a $d=3$ CFT takes the general form
\beq
\label{formee}
S_{EE} = a_1\frac{\ell}{\delta}-a(\theta)\;\mathrm{ln}\frac{\ell}{\delta}+O(1)
\eeq
where $\delta$ is a short-distance cutoff and $\ell$ is a length scale associated with the size of the subregion. The first term above is the area-law term, while the second term, which is universal, only appears in cases when the subregion has a sharp corner with opening angle $\theta$, henceforth referred to as the corner term. It has been conjectured based on holographic, free-field and numerical calculations, that in the smooth limit $\theta \to \pi$, the corner term in any $d=3$ CFT behaves as
\beq\label{corner}
a(\theta) = \frac{\pi^2 C_T}{24}(\theta -\pi)^2+\cdots
\eeq
where $C_T$, once again, is the coefficient appearing in the two-point function of stress tensors in that CFT. Here, we will show that our formula for the entanglement density directly reproduces equation \eqref{corner}. 

We start with our formula for the planar case in $d=3$:
\beq\label{3dED}
\delta^{(2)} S_{EE} = -\frac{\pi^2 C_T}{4}\int_{-\infty}^{\infty} dx_a\int_{-\infty}^{\infty} dx_b\;\frac{\chi(x_a)\chi(x_b)}{(x_a-x_b)^{4}}
\eeq
and consider the special shape deformation:
\beq
\label{cshape}
\chi(x) = \begin{cases} 0 & |x| > L \\ \frac{\alpha}{2L} (L^2-x^2) & | x| < L \end{cases}
\eeq
which has two sharp corners at $x=\pm L$ with opening angle $\theta = (\pi-\alpha)$. 
The two corners will both lead to independent logarithmic divergences which we can then isolate. 
We choose this form because it does not have any IR issues and because it is easy to work with analytically.\footnote{A somewhat similar form was used in \cite{Bueno:2015lza} to show that the Mezei formula reproduces the corner term.}

Note that in order to do the integrals in \eqref{3dED}  we are forced to confront UV divergences when the two-points $x_a$ and $x_b$ come together. This is then related to local contact terms in the entanglement density that we have so far avoided discussing. These terms are also related to the usual UV divergence of EE (that is the area law piece for $d=3$ shown in \eqref{formee}.)
An efficient way to deal with these contact terms is to use dimensional regularization
where the absence of a scale in the regulator means that we will only ever see logarithmic
divergences (which would then show up as $1/(d-3)$ poles.) Since we do not expect
logarithmic divergences in $d=3$ in the absence of sharp corners, this is then a good way of isolating the term of interest. To this end, we consider an entangling surface in a $d$-dimensional CFT
with the shape determined by \eqref{cshape} independent
of the other $d-3$ transverse directions. At second order the change in entanglement entropy is then:
\beq
\delta^{(2)} S_{EE} = -\frac{\pi^2 \alpha^2 C_T}{4(d+1)L^2}V_{d-3} \frac{\pi^{(d-3)/2} \Gamma(\frac{d+1}{2})} {\Gamma(d-1)}  \int_{-L}^{L}  dx_a\int_{-L}^{L} dx_b\;\frac{(L^2-x_a^2)(L^2-x_b^2)}{|x_a-x_b|^{d+1}}
\eeq
where $V_{d-3} = \mu^{d-3}$ is the volume of the transverse space.  This last
integral can easily be done (and converges for $d<0$) giving:
\beq
\delta^{(2)} S_{EE} = - \frac{\pi^2 \alpha^2 C_T(d-1) \Gamma(\frac{d-5}{2})}{2d(d+1)(d-2)\Gamma(d-1)} \left( \frac{\sqrt{\pi} \mu}{2L} \right)^{d-3}.
\eeq
Taking the limit $d\rightarrow 3$ we find the desired pole and logarithmic behavior: 
\beq
\delta^{(2)} S_{EE} = \frac{C_T \pi^2\alpha^2}{12}\left( \frac{1}{d-3}  - \log(2L/(\sqrt{\pi}\mu)) + \frac{\gamma}{2} - \frac{19}{12} + \mathcal{O}(d-3) \right).
\eeq
Since we had two corners with equal opening angles we can infer that:
\beq
a(\theta) = \frac{\pi^2C_T}{24}(\pi-\theta)^2
\eeq
which proves the conjecture of \cite{Bueno:2015rda, Bueno:2015xda}. We have thus shown that in $d=3$ the logarithmic corner term in the entanglement entropy is entirely captured by the non-local part of the entanglement density. Presumably similar remarks/proofs hold for higher dimensional cones as conjectured recently
in \cite{Bueno:2015lza} using the Mezei formula.

\subsection{Mezei formula}\label{Mezei}
Next we turn to proving the Mezei formula \cite{Mezei:2014zla} for the entanglement entropy across deformed spheres at second order in the shape deformation. Once again in this section, we will only be interested in spatial deformations. Let us expand the shape deformation $\chi(\Omega)$ in terms of real hyperspherical harmonics on the entangling surface $S^{d-2}$
\beq\label{defY}
\chi(\Omega) = \sum_{\ell,m_1,\cdots,m_{d-3}}a_{\ell,m_1,\cdots,m_{d-3}}Y_{\ell,m_1,\cdots,m_{d-3}}(\Omega).
\eeq
Based on holographic calculations in a large class of models, it was conjectured in \cite{Mezei:2014zla} that the universal contribution to the entanglement entropy at second order in the deformation is given by
\beq
S^{(2)}_{EE}= C_T\frac{\pi^{\frac{d+2}{2}}(d-1)}{2^{d-2}\Gamma(d+2)\Gamma(d/2)}\sum_{\ell,m_1,\cdots,m_{d-3}}a^2_{\ell,m_1,\cdots,m_{d-3}}\prod_{k=1}^d(\ell+k-2)\times \begin{cases} (-1)^{\frac{d-1}{2}}\frac{\pi}{2} & d\;\mathrm{odd} \\ (-1)^{\frac{d-2}{2}}\mathrm{ln}\,\frac{R}{\delta} & d\;\mathrm{even} \end{cases}.
\eeq
The positivity of the overall coefficient implies that the sphere is a local minimum for (the universal part of) entanglement entropy across all shapes with the same topology. Having derived the entanglement density for spheres from purely CFT considerations, we are now in a position to prove this conjecture. We start with our expression for the sphere entanglement density (setting $R=1$ for convenience)
\beq
S^{(2)}_{EE}=-\frac{\pi^2C_T}{(d+1)2^{d-1}}\int d^{d-2}\Omega_ad^{d-2}\Omega_b\;\chi(\Omega_a)\chi(\Omega_b)\frac{1}{(1-\Omega_a\cdot\Omega_b)^{d-1}}.
\eeq
Substituting equation \eqref{defY} , we obtain
\beq
S^{(2)}_{EE}=-\frac{\pi^2C_T}{(d+1)2^{d-1}}\sum_{\ell^a,m^a_1,\cdots,\ell^b,m^b_1,\cdots}a_{\ell^a,m^a_1,\cdots}a_{\ell^b,m^b_1,\cdots}\int d^{d-2}\Omega_ad^{d-2}\Omega_b\;\frac{Y_{\ell^a,m^a_1,\cdots,m^a_{d-3}}(\Omega_a)Y_{\ell^b,m^b_1,\cdots,m^b_{d-3}}(\Omega_b)}{(1-\Omega_a\cdot\Omega_b)^{d-1}}.
\eeq
From rotation invariance, it is evident that the integral above takes the form
\beq
\int d^{d-2}\Omega_ad^{d-2}\Omega_b\;\frac{Y_{\ell^a,m^a_1,\cdots,m^a_{d-3}}(\Omega_a)Y_{\ell^b,m^b_1,\cdots,m^b_{d-3}}(\Omega_b)}{(1-\Omega_a\cdot\Omega_b)^{d-1}}=c(\ell^a)\delta_{\ell^a\ell^b}\delta_{m_1^a,m_1^b}\cdots \delta_{m_{d-3}^a,m_{d-3}^b}.
\eeq
The constant on the right hand side can in turn be written as
\beq\label{c1}
c(\ell)=\frac{1}{\mathrm{dim}(H_{\ell})}\int d^{d-2}\Omega_ad^{d-2}\Omega_b\;\sum_{m_1,\cdots,m_{d-3}}\frac{Y_{\ell,m_1,\cdots,m_{d-3}}(\Omega_a)Y_{\ell,m_1,\cdots,m_{d-3}}(\Omega_b)}{(1-\Omega_a\cdot\Omega_b)^{d-1}}
\eeq
where $H_{\ell}$ is the space of all harmonics of order $\ell$. In order to explicitly compute $c(\ell)$, we can use the higher-dimensional analog of the addition theorem for spherical harmonics\footnote{We have normalized the hyperspherical harmonics as $\int_{S^{d-2}}d^{d-2}\Omega\;\left(Y_{\ell,m_1,\cdots,m_{d-3}}(\Omega)\right)^2 = 1$.}
\beq
\sum_{m_1,\cdots,m_{d-3}}Y_{\ell,m_1,\cdots,m_{d-3}}(\Omega_a)Y_{\ell,m_1,\cdots,m_{d-3}}(\Omega_b)=\frac{\mathrm{dim}(H_{\ell})}{vol(S^{d-2})}\frac{C^{(\frac{d-3}{2})}_{\ell}(\Omega_a\cdot \Omega_b)}{C^{(\frac{d-3}{2})}_{\ell}(1)}
\eeq
where $C^{(n)}_{\ell}(x)$ is the Gegenbauer polynomial. This allows us to perform all but one of the integrals in \eqref{c1} to obtain
\beq
c(\ell)=\lim_{z\to 1^+}\;\frac{\mathrm{vol}(S^{d-3})}{C^{(\frac{d-3}{2})}_{\ell}(1)}\int_0^{\pi}d\theta\;\sin^{d-3}\theta\frac{C^{(\frac{d-3}{2})}_{\ell}(\cos\theta)}{(z-\cos\theta)^{d-1}}.
\eeq
Note that we have also introduced the regulator $z$ above to control the divergences which arise in the $\theta\to 0$ (coincident) limit. Fortunately, there exists a closed form expression for the above $\theta$ integral \cite{Cohl, RS}
\beq
\int_0^{\pi}d\theta\;(\sin \theta)^{D-1}\frac{C^{\left(\frac{D-1}{2}\right)}_{\ell}(\cos\theta)}{(z - \cos\theta)^{D+1}}=\frac{e^{-i(D/2+1)\pi}\sqrt{\pi}\;\Gamma(D+\ell-1)}{2^{D/2-2}\Gamma(\ell+1) \Gamma(\frac{D-1}{2})\Gamma(D+1)}\frac{1}{(z^2-1)^\frac{D+2}{4}}Q^{D/2+1}_{D/2+\ell-1}(z)
\eeq
where the associated Legendre function $Q^{D/2+1}_{D/2+\ell-1}(z)$ is defined in terms of the hypergeometric function as
\beq
Q^{D/2+1}_{D/2+\ell-1}(z)=\frac{e^{i\pi(D/2+1)}\sqrt{\pi}\;\Gamma(D+\ell+1)}{2^{D/2+\ell}\Gamma(D/2+\ell+\frac{1}{2})}\frac{z^{1-\ell}}{(z^2-1)^{\frac{D+2}{4}}}\;_{2}F_{1}\left(\frac{\ell-1}{2},\frac{\ell}{2};\frac{D}{2}+\ell+\frac{1}{2};\frac{1}{z^2}\right).
\eeq
Using these expressions, we obtain\footnote{The careful reader might observe that the functional form (in $z$) appearing above is very closely related to the (deformed) Ryu-Takayanagi surface, with $\sin \Theta =\frac{1}{z}$ playing the role of the bulk coordinate defined in \cite{Mezei:2014zla}. This motivates the identification $\epsilon = \frac{z^2-1}{z^2} \sim \left(\frac{\delta}{R}\right)^2$.}
\beq\label{c3}
c(\ell)=\lim_{z\to 1^+}\;\frac{2\pi^{\frac{d}{2}}}{\Gamma(\frac{d-2}{2})}\frac{\Gamma(d+\ell-1)\Gamma(d-3)}{2^{d+\ell-4} \Gamma(\frac{d-3}{2})\Gamma(d-1)\Gamma(d/2+\ell-\frac{1}{2})}\frac{z^{1-\ell}}{(z^2-1)^{\frac{d}{2}}}\;_{2}F_{1}\left(\frac{\ell-1}{2},\frac{\ell}{2};\frac{d}{2}+\ell-\frac{1}{2};\frac{1}{z^2}\right).
\eeq
All that remains to be done is to take the limit $z\to 1^+$. Let us first consider $d$ odd; in this case the hypergeometric function behaves as
\beq
\;_{2}F_{1}\left(\frac{\ell-1}{2},\frac{\ell}{2};\frac{d}{2}+\ell-\frac{1}{2};1-\epsilon\right)=\Big(a_0+a_1\epsilon+a_2\epsilon^2+\cdots\Big)+\epsilon^{d/2}\Big(b_0+b_1\epsilon+b_2\epsilon^2+\cdots\Big)
\eeq
Going back to \eqref{c3}, we see that $c(\ell)$ is divergent in the limit $z\to 1^+$. However, these divergences, as before, are associated with the coincident limit $\Omega_a \to \Omega_b$. A proper treatment of these divergences would require knowledge of contact terms in the entanglement density, which we are not in control of. However, we can extract the universal (cutoff independent) term in \eqref{c3}, which comes from the $\epsilon^{d/2}$ term in the expansion of the hypergeometric function close to $z=1$, where the corresponding coeffcient $b_0$ is given by
\beq
b_0=(-1)^{\frac{d+1}{2}}\frac{\Gamma(\frac{d}{2}+\ell-\frac{1}{2})}{\Gamma(\frac{\ell}{2})\Gamma(\frac{\ell-1}{2})\Gamma(\frac{d}{2}+1)}\pi.
\eeq
For $d$ even, the expansion of the hypergeometric function contains a logarithmic term $\epsilon^{d/2}\mathrm{ln}\,\epsilon$ which then gives the universal contribution to the entanglement entropy, and whose coefficient is given by
\beq
\tilde{b}_0=(-1)^{\frac{d-2}{2}}\frac{\Gamma(\frac{d}{2}+\ell-\frac{1}{2})}{\Gamma(\frac{\ell}{2})\Gamma(\frac{\ell-1}{2})\Gamma(\frac{d}{2}+1)}.
\eeq
Finally, putting everything together and simplifying gives
\beq
S^{(2)}_{EE} =  C_T\frac{\pi^{\frac{d+2}{2}}(d-1)}{2^{d-2}\Gamma(d+2)\Gamma(d/2)}\sum_{\ell,m_1,\cdots,m_{d-3}}a^2_{\ell,m_1,\cdots,m_{d-3}}\prod_{k=1}^d(\ell+k-2)\times \begin{cases} (-1)^{\frac{d-1}{2}}\frac{\pi}{2} & d\;\mathrm{odd} \\ (-1)^{\frac{d-2}{2}}\mathrm{ln}\,\frac{R}{\delta} & d\;\mathrm{even} \end{cases}
\eeq
which is precisely the formula conjectured in \cite{Mezei:2014zla}. 

As mentioned previously, in \cite{Bueno:2015lza} the Mezei conjecture was used to compute the universal corner term contributions to entanglement entropy in higher dimensions. Since we have now explicitly proved the Mezei conjecture, this also completes the derivation of the higher dimensional corner terms in \cite{Bueno:2015lza}.

\section{Discussion}

We have presented a proof of the universality of the non-local part of entanglement density in any CFT
in any dimension. The form of the entanglement density is fixed by conformal invariance and the overall coefficient is determined by $C_T$. 
We have also shown that this universality fits into a triangle of results that have been the focus of  recent studies/conjectures on the shape dependence of CFT entanglement entropy and that we summarize in Fig.~\ref{fig:fig4}.
\myfig{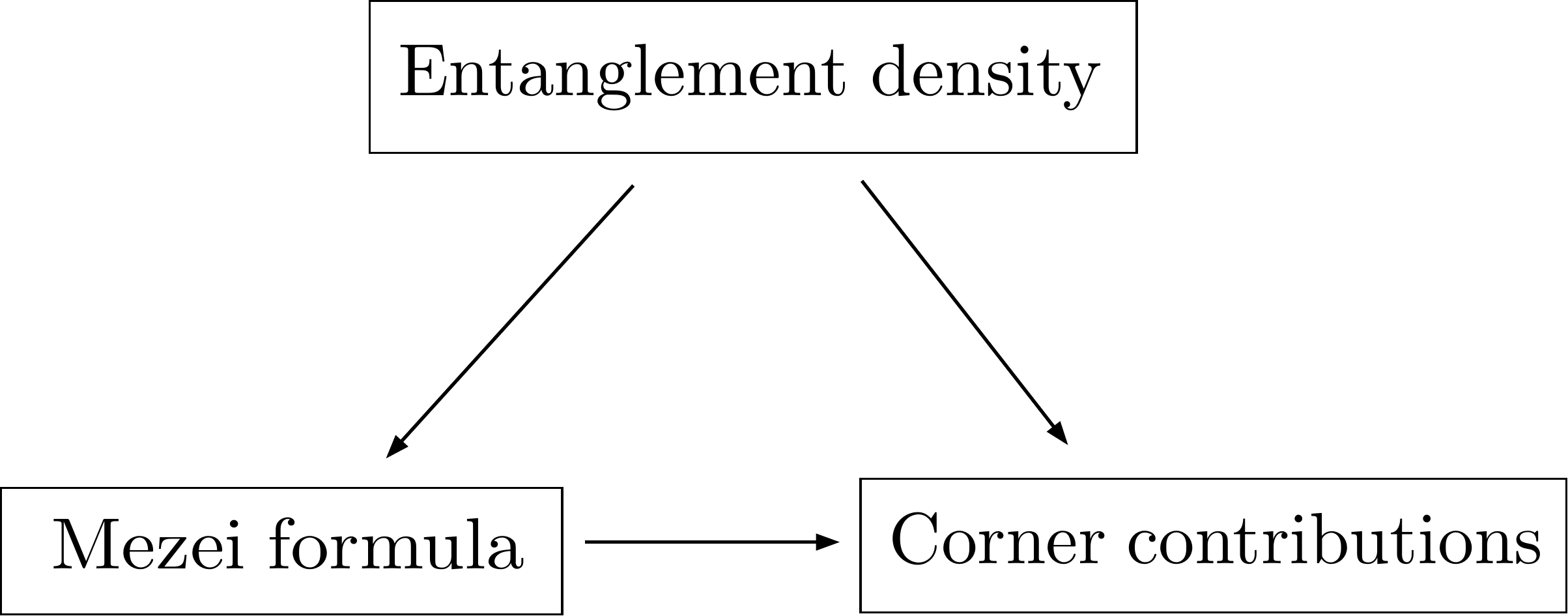}{9}{\textsf{The triangle of recent studies on the shape dependence of entanglement entropy in CFTs. The arrows denote implications.}}
We have a good understanding of the mechanism behind this universality.
The calculation presented here and previous works \cite{Rosenhaus:2014woa,Faulkner:2014jva} studied entanglement essentially in conformal perturbation theory, writing the answer in terms of $n$ point correlation functions on flat space up to some order $n$. Conformal invariance fixes the low point correlation functions that go into the calculation.  As we push these calculations to higher order in the  expansion parameter, four-point functions and higher  will appear and we expect universality to break down. At this point one might impose more restrictive conditions on
the CFT that are expected of a theory with a gravitational dual --- large-$N$ factorization and an appropriate sparseness condition on the low-lying spectrum of operator dimensions \cite{Heemskerk:2009pn} --- after which we would expect universality to re-emerge. Ryu-Takayangi taught us the surprising result that all CFTs with classical gravity duals have the same vacuum entanglement structure. We are far from being able to prove such a statement\footnote{For 2d CFTs however this statement has for the most part been established \cite{Hartman:2013mia}.}, but these calculations represent a first step. 
 
It is interesting to note that even before we reach the stage where universality breaks down, entanglement in this perturbative framework already displays rich features that are expected of a CFT with a holographic dual \cite{Faulkner:2014jva}. The relative entropy contribution studied in Section~\ref{sec:rel} arises from an integral over an operator located inside the domain of dependence of $A$, which reminds us of smearing functions that are used to construct local bulk fields \cite{Hamilton:2006az}. It is this feature of the CFT calculation that we think is responsible for probing deep into the bulk of an emergent AdS dual to measure the change in area of the minimal surface. 

We now discuss some possible generalizations that one may pursue.
Firstly it would be of interest to study the Renyi generalization of entanglement density (see \cite{Bianchi:2015liz} for related discussion). 
For example using the mutual information definition of entanglement density  \eqref{cmi},
we can imagine simply generalizing this by taking $I \rightarrow I_n$ where $I_n$ is called the Renyi
mutual information.  This captures the non-local part of the Renyi entanglement density which is then a UV finite quantity.
It is not hard to argue, based on conformal invariance alone, that Renyi entanglement density should take the same form as regular entanglement density:
\beq
\label{renyi}
S^{n\, (2)}_{\ui\uj,non-local}(x_a,x_b) = -e_n \frac{\eta_{\ui\uj}}{|x_a-x_b|^{2(d-1)}} 
\eeq
for some unfixed coefficient $e_n$ depending on the Renyi index $n$. As a function of $n$ we expect that  $e_n$ is non-universal and theory dependent. 
We do know that as $n\rightarrow 1$ it should equal $e_1 = 2\pi^2 C_T/(d+1)$.   In $d=3$ we can use this
result to make a prediction for the logarithmic term in the Renyi entropies in the presence
of a corner. Following the same steps as in Section~\ref{sec:corner} we find the coefficient of the log
is
\beq
a_n(\theta) = \frac{e_n^{(d=3)}}{12}(\pi-\theta)^2 + \ldots
\eeq
for opening angles close to $\pi$. Quite a bit is known about such contributions to the corner terms in Renyi entropies which we may then use to make predictions about $e_n$. For example
the conjecture in \cite{Bueno:2015qya} would lead to the relation:
\beq
e_n = \frac{ 12 h_n}{(n-1)\pi}
\eeq
where $h_n$ is  the (higher dimensional) twist operator dimension which can be related to a one-point function of the stress tensor for the CFT living on the space $\mathbb{H}_{d-1} \times S^1_{(n)}$ where the radius of the circle $S^1$ has been enlarged by a factor of $n$ relative
to the conformally flat version.   Thus if we could establish \eqref{renyi} for the entanglement density then it would
prove the conjecture of \cite{Bueno:2015qya}. Our field theory approach applied to this problem would naively suggest $e_n$ is related to an integrated (connected) stress tensor  two-point function for the CFT
living on $\mathbb{H}_{d-1} \times S^1_{(n)}$ and it is not at all clear (without putting too much thought into it) how this could be related to a one-point function on the same space. 

Further generalizations, that we hope to pursue in the future, include the computation of higher order terms in the perturbative expansion of the shape deformation as well as studying entanglement density in relativistic non-conformal theories. 

Finally we comment on previous studies of second order shape deformations of EE in \cite{Rosenhaus:2014zza} using
similar CFT techniques (see also \cite{Lewkowycz:2014jia} for the Renyi entropy case). These authors considered
more general metric deformations that contain the shape deformation as a special case and attempt to access the universal  logarithmic divergences in EE for $d=4$ - first written down using different arguments in \cite{Solodukhin:2008dh}. This should be contrasted with our approach of examining the non-local (finite) shape dependent part of EE. Certain issues with this CFT perturbative approach were identified in \cite{Rosenhaus:2014zza}, which we suspect would be resolved by a more careful analysis of the relative entropy term along the lines in this paper. However it is not clear that these universal $ln$ terms can be extracted using a non-local finite term in EE, say deformed by a more general metric, which was the origin of the many simplifications that occurred in our calculation.

\subsection*{Acknowledgements}
We would like to thank Srivatsan Balakrishnan, Ling-Yan Hung, Mark Mezei, Rob Myers, Mukund Rangamani, Vladimir Rosenhaus, Misha Smolkin and Huajia Wang for helpful conversations. TF and OP would like to thank the Kavli Institute for Theoretical Physics, Santa Barbara where some of this work was carried out, for support under the U.S. National Foundation grant number NSF PHY11-25915. Work supported in part by the U.S. Department of Energy contract DE-FG02-13ER42001. TF is supported by the DARPA, YFA
Grant  No. D15AP00108. 

\appendix 
\numberwithin{equation}{section}
\section{Perturbative change in the reduced density matrix}\label{appA}
In this appendix, we prove the formula \eqref{PEDM} for the perturbative expansion of the reduced density matrix on the subregion $A$ in terms of the metric perturbation $\delta g_{\mu\nu}$. Consider a generic relativistic quantum field theory that admits a path-integral description in terms of the field $\phi$, which collectively denotes all the fields over which we integrate. The density matrix corresponding to the ground state wavefunction on the Cauchy surface $\Sigma$ is given by
\beq
|0\rangle\langle 0| = \mathrm{lim}_{\beta \to \infty}\;\frac{e^{-\beta \widehat{H}}}{\mathrm{tr}_{\mathfrak{h}_{\Sigma}}\left(e^{-\beta \widehat{H}}\right)}
\eeq
where $\widehat{H}$ is the Hamiltonian, and $\mathfrak{h}_{\Sigma}$ is the entire Hilbert space. In the path integral language, we can describe a matrix element of this density matrix as a product of path integrals over the regions $x^0_E>0$ and $x^0_E<0$ of Euclidean space $\re^d$ with the appropriate boundary conditions
\beq
\langle \phi_-|0\rangle\langle 0|\phi_+\rangle = \frac{1}{Z}\int_{\phi(0^-,\bdx)= \phi_-(\bdx)}[D\phi]_{x^0_E<0}\;e^{-S[\phi]}\int_{\phi(0^+,\bdx)= \phi_+(\bdx)}[D\phi]_{x^0_E>0}\;e^{-S[\phi]}
\eeq
where we have denoted the spatial coordinates collectively as $\bdx=(x^1,x^i)$, and $Z$ is the partition function of the theory on $\re^d$. 

Let us now denote the reduced density matrix for the half space $A=\{\bdx_A=(x^1,x^i)|x^1>0\}$ by $\brho_0$. The matrix element $\langle \phi^A_-|\brho_0|\phi^A_+\rangle$ is then given by gluing the above path integrals along the complementary space $\bar{A}$ at $x^0_E=0$ 
\beq
\langle \phi^A_-|\brho_0|\phi^A_+\rangle=\frac{1}{Z}\int^{\phi(0^-,\bdx_A)=\phi^A_-}_{\phi(0^+,\bdx_A)=\phi^A_+}[D\phi]\;e^{-S[\phi]}
\eeq
where $\bdx_A$ are spatial coordinates on $A$, and $\phi^A_{\pm}$ denotes a field configuration restricted to $A$. By slicing this path integral along the angular direction $\theta$ in the $(x^0_E,x^1)$ plane, it becomes immediately clear that the reduced density matrix can be written in operator form as \cite{Bisognano:1976za}
\beq\label{EH}
\brho_0 = \frac{e^{-2\pi\widehat{K}}}{\mathrm{tr}_{\mathfrak{h}_A}\left(e^{-2\pi\widehat{K}}\right)}
\eeq
where $\mathfrak{h}_A$ is the Hilbert space on $A$, and $\widehat{K}$ is the generator of $\theta$-rotations 
\beq
\hK = \int d^{d-2}x^i\int_0^{\infty}dx^1\;x^1\;\widehat{T}^{00}(0,x^1,x^i).
\eeq
From equation \eqref{EH}, we see that up to an overall shift coming from the normalization, the entanglement Hamiltonian in this case is given by 
\beq
\widehat{H}_E \equiv -\mathrm{ln}\;\brho_0=2\pi\widehat{K}+\mathrm{constant}.
\eeq

Next, we turn on a small (background) metric deformation $\delta g_{\mu\nu}$. The new reduced density matrix $\brho$ is given by
\beq
\langle \phi^A_-|\brho|\phi^A_+\rangle=\frac{1}{(Z+\delta Z)}\int^{\phi(0^-,\bdx^A)=\phi^A_-}_{\phi(0^+,\bdx^A)=\phi^A_+}[D\phi]\;e^{-S[\phi]+\frac{1}{2}\int d^d\bx\;\delta g^{\mu\nu}(\bx)T_{\mu\nu}(\bx)+\cdots}\label{dm1}
\eeq
where we have introduced $\bx^{\mu}=(x^0_E,\bdx)$ to collectively denote the coordinates on $\re^d$. One might worry about an extra term in the exponential coming from a change in the stress tensor upon introducing $\delta g$. However, at second order in the $\delta g$-expansion, such a term can at best give a local contribution, and so we drop it. Therefore to quadratic order in $\delta g$, we have
\beqn\label{PEDM0a}
\langle \phi^A_-|\delta\brho|\phi^A_+\rangle &=& \frac{1}{2Z}\int_{\phi^A_+}^{\phi_-^A}[D\phi]\;e^{-S[\phi]}\Big\{\frac{1}{2}\int d^{d}\bx\;\delta g_{\mu\nu}(\bx)\;\Big(T^{\mu\nu}(\bx)-\langle T^{\mu\nu}(\bx)\rangle\Big)\\
&+&\frac{1}{8}\int  d^{d}\bx_a d^{d}\bx_b\;\delta g_{\mu\nu}(\bx_a)\delta g_{\lambda\sigma}(\bx_b) \;\Big(T^{\mu\nu}(\bx_a)T^{\lambda\sigma}(\bx_b)-2T^{\mu\nu}(\bx_a)\;\left\langle T^{\lambda\sigma}(\bx_b)\right\rangle\nonumber\\
&-&\left\langle T^{\mu\nu}(\bx_a)T^{\lambda\sigma}(\bx_b)\right\rangle+2\left\langle T^{\mu\nu}(\bx_a)\right\rangle\;\left\langle T^{\lambda\sigma}(\bx_b)\right\rangle\Big)+\cdots\Big\}.
\eeqn
From equation \eqref{EH}, we can then infer the following operator expression for the change in the reduced density matrix
\beqn\label{PEDM0b}
\delta\brho &=& \frac{1}{2}\int d^{d}\bx\;\delta g_{\mu\nu}(\bx)\;\brho_0\Big(\hT^{\mu\nu}(\bx)-\mathrm{tr}_{\mathfrak{h}_A}\left(\brho_0\hT^{\mu\nu}(\bx)\right)\Big)\\
&+&\frac{1}{8}\int  d^{d}\bx_a d^{d}\bx_b\;\delta g_{\mu\nu}(\bx_a)\delta g_{\lambda\sigma}(\bx_b) \;\brho_0\Big\{\mathcal{T}\left[\hT^{\mu\nu}(\bx_a)\hT^{\lambda\sigma}(\bx_b)\right]-2\hT^{\mu\nu}(\bx_a)\;\mathrm{tr}_{\mathfrak{h}_A}\left(\brho_0\hT^{\lambda\sigma}(\bx_b)\right)\nonumber\\
&-&\mathrm{tr}_{\mathfrak{h}_A}\left(\brho_0\mathcal{T} \left[\hT^{\mu\nu}(\bx_a)\hT^{\lambda\sigma}(\bx_b)\right]\right)+2\mathrm{tr}_{\mathfrak{h}_A}\left(\brho_0\hT^{\mu\nu}(\bx_a)\right)\;\mathrm{tr}_{\mathfrak{h}_A}\left(\brho_0\hT^{\lambda\sigma}(\bx_b)\right)\Big\}+\cdots.
\eeqn
Switching to polar coordinates $(r,\theta)$ in the $(x_0^E,x^1)$ plane, the operator $\hT(\theta, r,x^i)$ above is to be interpreted (from the point of view of the reduced density matrix) as 
\beq
\hT^{\mu\nu}(\theta,r,x^i) = {\left(R(\theta)\right)^{\mu}}_{\lambda}{\left(R(\theta)\right)^{\nu}}_{\sigma}e^{\theta \hK}\hT^{\lambda\sigma}(0, r, x^i)e^{-\theta \hK}
\eeq
where $R(\theta)$ is the appropriate rotation matrix in the vector representation. Further,  $\mathcal{T}$ is the angular-ordering operator in the $(x_E^0,x^1)$ plane
\beq
\mathcal{T}\left[ \widehat{\mathcal{O}}(\theta_a)\widehat{\mathcal{O}}(\theta_b)\right] = \widehat{\mathcal{O}}(\theta_a)\widehat{\mathcal{O}}(\theta_b) H (\theta_a-\theta_b)+\widehat{\mathcal{O}}(\theta_b)\widehat{\mathcal{O}}(\theta_a)H (\theta_b-\theta_a)
\eeq
where $H (\theta_a-\theta_b)$ is the Heaviside step function.

\section{Angular ordering in the relative entropy term}
Recall that the relative entropy term is given by
\beq
\delta S^{(2)}_{EE}=-\frac{1}{4}\int d^d\bx_a d^d\bx_b\;\delta g_{\mu\nu}(\bx_a)\delta g_{\lambda\sigma}(\bx_b)\int_{0}^{\infty}d\beta\beta\;\mathrm{tr}_{\mathfrak{h}_A}\Big(\frac{\brho_0}{(\brho_0+\beta)^2}:\hT^{\mu\nu}:(\bx_a)\frac{\brho_0}{\brho_0+\beta}:\hT^{\lambda\sigma}:(\bx_b)\Big)
\eeq
where we have used the short notation
\beq
:\hT^{\mu\nu}:(\bx_a)=\hT^{\mu\nu}(\bx_a)-\mathrm{tr}_{\mathfrak{h}_A}\left(\brho_0\;\hT^{\mu\nu}(\bx_a)\right).
\eeq
Unfortunately, the above expression is not $\mathcal{T}$-ordered, and cannot be written in terms of a Euclidean correlation function. To resolve this problem, we will perform the $\beta$-integral, and then manipulate the expression further to bring it into a $\mathcal{T}$-ordered form. Let's begin by rewriting it as
\beqn\label{re1}
\delta S^{(2)}_{EE}&=&-\frac{1}{4}\int d^d\bx_a d^d\bx_b\;\widetilde{\delta g}_{\mu\nu}(\bx_a)\widetilde{\delta g}_{\lambda\sigma}(\bx_b)\\
&\times &\int_{0}^{\infty}d\beta\beta\;\mathrm{tr}_{\mathfrak{h}_A}\Big(\frac{\brho_0}{(\brho_0+\beta)^2}e^{\theta_a \hK}:\hT^{\mu\nu}:(0,r_a,x_a^i)e^{-\theta_a \hK}\frac{\brho_0}{\brho_0+\beta}e^{\theta_b \hK}:\hT^{\lambda\sigma}:(0,r_b,x^i_b)e^{-\theta_b \hK}\Big)\nonumber
\eeqn
where we have defined
\beq
\widetilde{\delta g}_{\mu\nu}(\bx) = \delta g_{\lambda\sigma}(\bx){\left(R(\theta)\right)^{\lambda}}_{\mu}{\left(R(\theta)\right)^{\sigma}}_{\nu}.
\eeq
Let us denote the first line of \eqref{re1} as $-\int d\tilde{\mu}_{\mu\nu\lambda\sigma}$ for convenience. Using the eigenstates of $\hK$ defined by $\hK|\omega\rangle = \omega |\omega\rangle$ to carry out the above trace, and writing $\brho_0=ce^{-2\pi\hK}$, we get
\beqn\label{exp1}
\delta S^{(2)}_{EE}&=&-\int d\tilde{\mu}_{\mu\nu\lambda\sigma}\sum_{\omega_a,\omega_b}\; e^{(\theta_a-\theta_b)(\omega_a-\omega_b)}\langle\omega_a|:\hT^{\mu\nu}:(0,r_a,x^i_a)|\omega_b\rangle\langle\omega_b|:\hT^{\lambda\sigma}:(0,r_b,x^i_b)|\omega_a\rangle\nonumber\\
&\times& c\int_0^{\infty}d\beta\beta \frac{e^{-2\pi(\omega_a+\omega_b)}}{(e^{-2\pi\omega_a}+\beta)^2(\beta+e^{-2\pi\omega_b)}}.
\eeqn
The $\beta$ integral can be performed to obtain
\beq
\int_0^{\infty}d\beta\beta \frac{e^{-2\pi(\omega_a+\omega_b)}}{(e^{-2\pi\omega_a}+\beta)^2(\beta+e^{-2\pi\omega_b})}=e^{-2\pi\omega_b}\left(\frac{\nu e^{\nu}}{(1-e^{\nu})^2}+\frac{1}{1-e^{\nu}}\right)
\eeq
where $\nu = 2\pi(\omega_a-\omega_b)$. Next, using the formulae
\beq\label{FT1}
\left(\frac{1}{1-e^{\nu}}+\frac{\nu e^{\nu}}{(1-e^{\nu})^2}\right)=\int^{\infty-i\varepsilon}_{-\infty-i\varepsilon}\frac{ds}{2\pi i}e^{-i\nu s/2\pi}\frac{s}{4\sinh^2(s/2)}
\eeq
\beq\label{FT2}
\left(\frac{1}{1-e^{\nu}}+\frac{\nu e^{\nu}}{(1-e^{\nu})^2}\right)=\int^{\infty+i\varepsilon}_{-\infty+i\varepsilon}\frac{ds}{2\pi i}e^{-i\nu s/2\pi-\nu}\frac{s-2\pi i}{4\sinh^2(s/2)}
\eeq
allows us to revert back from the spectral representation to the operator-trace form. To proceed, let's split the integral in eq \eqref{exp1} into two parts: $\theta_a>\theta_b$ and $\theta_b>\theta_a$. For the first integral, we use equation \eqref{FT2} and for the second integral we use \eqref{FT1}
\beqn\label{re2}
\delta S_{EE,\theta_a >\theta_b}^{(2)} &=& \int_{\theta_a>\theta_b} d\mu_{\mu\nu\lambda\sigma}\int_{\re+i\varepsilon}\frac{ds}{2\pi i}\\
&\times &\;\frac{2\pi i-s}{4\mathrm{sinh}^2(s/2)}{\left(R^{-1}(is)\right)^{\lambda}}_{\kappa}{\left(R^{-1}(is)\right)^{\sigma}}_{\eta}\mathrm{tr}_{\mathfrak{h}_A,\mathrm{conn.}}\Big(\brho_0\hT^{\mu\nu}(\theta_a,r_a,x^i_a)\hT^{\kappa\eta}(\theta_b+is,r_b,x^i_b)\Big)\nonumber
\eeqn
\beqn\label{re3}
\delta S_{EE,\theta_a <\theta_b}^{(2)} &=& -\int_{\theta_a<\theta_b} d\mu_{\mu\nu\lambda\sigma}\int_{\re-i\varepsilon}\frac{ds}{2\pi i}\\
&\times &\frac{s}{4\mathrm{sinh}^2(s/2)}\;{\left(R^{-1}(is)\right)^{\lambda}}_{\kappa}{\left(R^{-1}(is)\right)^{\sigma}}_{\eta}\mathrm{tr}_{\mathfrak{h}_A,\mathrm{conn.}}\Big(\brho_0\hT^{\kappa\eta}(\theta_b+is,r_b,x^i_b)\hT^{\mu\nu}(\theta_a,r_a,x^i_a)\Big)\nonumber
\eeqn
where $\int d\mu_{\mu\nu\lambda\sigma}=\frac{1}{4}\int d^d\bx_a d^d\bx_b\;\delta g_{\mu\nu}(\bx_a)\delta g_{\lambda\sigma}(\bx_b)$, and we have introduced the connected trace
\beq
\mathrm{tr}_{\mathfrak{h}_A,\mathrm{conn.}}\left(\brho_0\widehat{A}\cdot\widehat{B}\right) =\mathrm{tr}_{\mathfrak{h}_A}\left(\brho_0\widehat{A}\cdot\widehat{B}\right) - \mathrm{tr}_{\mathfrak{h}_A}\left(\brho_0\widehat{A}\right) \mathrm{tr}_{\mathfrak{h}_A}\left(\brho_0\widehat{B}\right).
\eeq
Now making the replacements $\bx_a\leftrightarrow \bx_b$ and $s\to -s$ in \eqref{re3}, and adding \eqref{re2} and \eqref{re3}, we obtain 
\beq
\delta S_{EE}^{(2)} = \int_{\theta_a>\theta_b} d\mu_{\mu\nu\lambda\sigma}\int_{+i\varepsilon}\frac{ds}{4\mathrm{sinh}^2(s/2)}\;{\left(R^{-1}(is)\right)^{\lambda}}_{\kappa}{\left(R^{-1}(is)\right)^{\sigma}}_{\eta}\mathrm{tr}_{\mathfrak{h}_A,\mathrm{conn.}}\Big(\brho_0\hT^{\mu\nu}(\theta_a,r_a,x^i_a)\hT^{\kappa\eta}(\theta_b+is,r_b,x^i_b)\Big).
\eeq
Finally, once again making the replacements $\bx_a\leftrightarrow \bx_b$ and $s\to -s$ in the above integral, and adding to itself, we obtain
\beqn
\delta S^{(2)}_{EE} &=& \frac{1}{8}\int\;d^d\bx_a d^d\bx_b\;\delta g_{\mu\nu}(\bx_a)\delta g_{\lambda\sigma}(\bx_b)\\
&\times &\int_{C}\frac{ds}{4\mathrm{sinh}^2(s/2)}\;{\left(R^{-1}(is)\right)^{\lambda}}_{\kappa}{\left(R^{-1}(is)\right)^{\sigma}}_{\eta}\mathrm{tr}_{\mathfrak{h}_A,\mathrm{conn.}}\Big(\brho_0\mathcal{T}[\hT^{\mu\nu}(\theta_a,r_a,x^i_a)\hT^{\kappa\eta}(\theta_b+is,r_b,x^i_b)]\Big)\nonumber
\eeqn
where the contour is given by $C = \re+i\varepsilon \;\mathrm{sign}(\theta_a-\theta_b)$. This then gives the result \eqref{relent1} used in the main text. 

\section{Contact \&\ Crossing terms}
In this appendix, we analyse (i) the contact terms in equation \eqref{mH0}, (ii) crossing terms in the three-point function term in \eqref{mH0}.
\subsection{Contact terms}
The contact terms in \eqref{mH0} are given by
\beqn\label{ct1}
\mathrm{contact\;terms}&=&\int_{\mathcal{H}_{\Lambda}}\int_{\pa \mathcal{H}_{\Lambda}}\xi_{\beta}(y_a)n_{\gamma}(y_b)\xi_{\delta}(y_b)\nabla^{(a)}_{\alpha}\left\langle \hT^{\alpha\beta}(y_a)\hT^{\gamma\delta}(y_b)\widehat{H}_{E}\right\rangle\nonumber\\
&+&\frac{1}{2}\int_{\mathcal{H}_{\Lambda}}\int_{\mathcal{H}_{\Lambda}}\xi_{\beta}(y_a)\xi_{\delta}(y_b)\nabla^{(a)}_{\alpha}\nabla_{\gamma}^{(b)}\left\langle \hT^{\alpha\beta}(y_a)\hT^{\gamma\delta}(y_b)\widehat{H}_{E}\right\rangle\nonumber\\
&+& \frac{1}{2}\int_{\pa \mathcal{H}_{\Lambda}}\int_{\mathcal{H}_{\Lambda}}\;n_{\alpha}(y_a)\xi_{\beta}(y_a)\Xi(y_b)\left\langle \hT^{\alpha\beta}(y_a){\hT^{\gamma}}_{\gamma}(y_b)\widehat{H}_{E}\right\rangle\nonumber\\
&-&\frac{1}{2}\int_{\mathcal{H}_{\Lambda}}\int_{\mathcal{H}_{\Lambda}}\;\xi_{\beta}(y_a)\Xi(y_b)\nabla_{\alpha}^{(a)}\left\langle \hT^{\alpha\beta}(y_a){\hT^{\gamma}}_{\gamma}(y_b)\widehat{H}_{E}\right\rangle\nonumber\\
&+&\frac{1}{8} \int_{\mathcal{H}_{\Lambda}}\int_{\mathcal{H}_{\Lambda}}\; \Xi(y_a)\Xi(y_b)\left\langle {\hT^{\alpha}}_{\alpha}(y_a){\hT^{\gamma}}_{\gamma}(y_b)\widehat{H}_E\right\rangle
\eeqn
where we have defined $\Xi = \xi^{\alpha}\pa_{\alpha}\mathrm{ln}\;\Omega^2$. The modular Hamiltonian can be written as the following integral over the constant $\tau=\tau_c$ slice 
\beq
\widehat{H}_E=2\pi\int_{\mathbb{H}^{d-1}}\frac{d^{d-2}x^i_cdz_c}{z_c^{d-1}}\;\hT_{\tau\tau}(\tau_c,z_c,x^i_c)+\mathrm{constant}.
\eeq
We will need to use the diffeomorphism and trace Ward identities for three-point functions of stress tensors, which can be found in \cite{Osborn:1993cr} (for Euclidean space). On a general manifold with the metric $g_{\mu\nu}$, these Ward identities take the following form 
\beqn\label{DWI}
\nabla_{(x)}^{\mu}\left\langle \hT_{\mu\nu}(x)\hT_{\lambda\sigma}(y)\hT_{\rho\kappa}(z)\right\rangle &=& \nabla_{\nu}^{(x)}\left(\frac{\delta^d(x-y)}{\sqrt{g}}\right)\left\langle \hT_{\lambda\sigma}(x)\hT_{\rho\kappa}(z)\right\rangle\nonumber\\
&+&2\nabla^{(x)}_{(\lambda}\left(\frac{\delta^d(x-y)}{\sqrt{g}}\left\langle \hT_{\sigma)\nu}(x)\hT_{\rho\kappa}(z)\right\rangle\right)\nonumber\\
&+& \nabla^{(x)}_{\nu}\left(\frac{\delta^d(x-z)}{\sqrt{g}}\right)\left\langle \hT_{\lambda\sigma}(y)\hT_{\rho\kappa}(x)\right\rangle\nonumber\\
&+&2\nabla_{(\rho}^{(x)}\left(\frac{\delta^d(x-z)}{\sqrt{g}}\left\langle \hT_{\kappa)\nu}(x)\hT_{\lambda\sigma}(y)\right\rangle\right)
\eeqn
\beq\label{TWI}
g^{\mu\nu}(x)\left\langle \hT_{\mu\nu}(x)\hT_{\lambda\sigma}(y)\hT_{\rho\kappa}(z)\right\rangle = 2\left(\frac{\delta^d(x-y)}{\sqrt{g}}+\frac{\delta^d(x-z)}{\sqrt{g}}\right)\left\langle \hT_{\lambda\sigma}(y)\hT_{\rho\kappa}(z)\right\rangle.
\eeq
Using these identities, we see that most of the contact terms in \eqref{ct1} drop out trivially because $y_a$ and $y_b$ are well separated (i.e., because we are only keeping terms which contribute to the non-local part of the entanglement density). The only potentially non-trivial terms are
\beqn\label{ct2}
\mathrm{contact\;terms}&=&-\int_{\mathcal{H}_{\Lambda}}\int_{\pa \mathcal{H}_{\Lambda}}\xi_{\beta}(y_a)n_{\gamma}(y_b)\xi_{\delta}(y_b)\nabla^{(a)}_{\alpha}\left\langle \hT^{\alpha\beta}(y_a)\hT^{\gamma\delta}(y_b)\widehat{H}_{E}\right\rangle\nonumber\\
&+& \frac{1}{2}\int_{\pa \mathcal{H}_{\Lambda}}\int_{\mathcal{H}_{\Lambda}}\;n_{\alpha}(y_a)\xi_{\beta}(y_a)\Xi(y_b)\left\langle \hT^{\alpha\beta}(y_a){\hT^{\gamma}}_{\gamma}(y_b)\widehat{H}_{E}\right\rangle.
\eeqn
Before proceeding to consider these terms, we first pause to observe that both the above terms are independent of the time coordinate $\tau_c$ at which the modular Hamiltonian is placed. This is because $\widehat{H}_E$ is a conserved charge, and we can freely move it in $\tau_c$ as long as we don't cross other operators. When we do in fact cross another operator, say $\hT^{\alpha\beta}(y_a)$, then using the commutator $\left[\widehat{H}_E, \hT^{\alpha\beta}(y_a)\right] \sim \pa_{\tau_a}\hT^{\alpha\beta}(y_a)$, we generate extra terms involving the two-point functions of stress tensors. However, using the Ward identities for the 2-point functions, it is straightforward to check that such crossing terms in \eqref{ct2} vanish for $y_a$ and $y_b$ well-separated. Thus, we conclude that both the terms in \eqref{ct2} are independent of $\tau_c$.  

In light of the above discussion, we can simplify the integrals in \eqref{ct2} by integrating over $\tau_c $ (and dividing by $2\pi$). For instance, the first term in \eqref{ct2} upon using the diffeomorphism Ward identity and integrating over $\tau_c$ gives
\beqn
1st\;\mathrm{term} &=& -\frac{1}{2\pi}\int_{\mathcal{H}_{\Lambda}}d\mu_a\int_{\pa \mathcal{H}_{\Lambda}}d\bar{\mu}_b\int_{\mathcal{H}_{\Lambda}}d\mu_c\;\xi^{\beta}(y_a)n^{\gamma}(y_b)\xi^{\delta}(y_b)\\
&\times &\left[\nabla_{\beta}^{(a)}\left(\frac{\delta(y_a-y_c)}{\sqrt{g^{\mathcal{H}}}}\right)\left\langle \hT_{\gamma\delta}(y_b)\hT_{\tau\tau}(y_a)\right\rangle+ 2\partial^{(a)}_{\tau}\left(\frac{\delta(y_a-y_c)}{\sqrt{g^{\mathcal{H}}}}\left\langle \hT_{\beta \tau}(y_a)\hT_{\gamma\delta}(y_b)\right\rangle\right)\right]\nonumber\\
&=& \frac{1}{\pi}\int_{\mathcal{H}_{\Lambda}} d\mu_a\int_{\pa \mathcal{H}_{\Lambda}}d\bar{\mu}_b\;\partial^{(a)}_{\tau}\xi^{\beta}(y_a)n^{\gamma}(y_b)\xi^{\delta}(y_b)\;\left\langle \hT_{\beta \tau}(y_a)\hT_{\gamma\delta}(y_b)\right\rangle
\eeqn
where in the second line we have performed the $y_c$ integration, and once again we have taken $y_a$ and $y_b$ to be well-separated. The two-point function appearing above can be computed efficiently using the embedding space formalism. Having done so, one finds that the above term is suppressed by a factor of $\frac{1}{\Lambda}$. The only thing to check is whether the $z_a$ integral inside $d\mu_a$ is divergent, because such divergences could give potential enhancements. Happily, one finds that the $z_a$ integral is finite, and thus the above term vanishes in the limit $\Lambda \to \infty.$ Similarly, one can check that the second term in \eqref{ct2} also vanishes as $\Lambda \to \infty.$

\subsection{Crossing terms}
Next, we argue that the three-point function term in \eqref{mH0} 
\beq\label{mHapp}
\delta S_{EE}^{(1)}=\frac{1}{8}\int_{\pa\mathcal{H}_{\Lambda}} d\bar{\mu}_a n^{\alpha}(y_a)\xi^{\beta}(y_a) \int_{\pa\mathcal{H}_{\Lambda}} d\bar{\mu}_b n^{\gamma}(y_b)\xi^{\delta}(y_b)\;\left\langle \hT_{\alpha\beta}(y_a)\hT_{\gamma\delta}(y_b)\widehat{H}_E\right\rangle_{\mathcal{H}}
\eeq
is independent of the time $\tau_c $ at which we place the modular Hamiltonian. Since the modular Hamiltonian is a conserved charge, we are indeed free to move it around in $\tau$, as long as we don't cross another operator insertion. However, when we do cross another operator, we pick up an extra contact (or commutator) term, which we will refer to as a \emph{crossing term}. For instance, let us take $\tau_c$ from $\tau_a-\epsilon$ to $\tau_a+\epsilon$; in this case we pick up the crossing term
\beqn
&= &\frac{1}{8}\int_{\pa\mathcal{H}_{\Lambda}} d\bar{\mu}_a n^{\alpha}(y_a)\xi^{\beta}(y_a) \int_{\pa\mathcal{H}_{\Lambda}} d\bar{\mu}_b n^{\gamma}(y_b)\xi^{\delta}(y_b)\;\left\langle \left[\widehat{H}_E, \hT_{\alpha\beta}(y_a)\right]\hT_{\gamma\delta}(y_b)\right\rangle_{\mathcal{H}}\\
&= &\frac{1}{8}\int_{\pa\mathcal{H}_{\Lambda}} d\bar{\mu}_a n^{\alpha}(y_a)\xi^{\beta}(y_a) \int_{\pa\mathcal{H}_{\Lambda}} d\bar{\mu}_b n^{\gamma}(y_b)\xi^{\delta}(y_b)\;\pa_{\tau_a}\left\langle \hT_{\alpha\beta}(y_a)\hT_{\gamma\delta}(y_b)\right\rangle_{\mathcal{H}}.
\eeqn
Similar to our previous discussion, the two-point function appearing above can be computed using the embedding space formalism. Having done so, one finds that the above term is suppressed by a factor of $\frac{1}{\Lambda^2}$. There are no other enhancements to cancel this factor, and the above term simply vanishes in the limit $\Lambda \to \infty.$ Therefore in this limit, the crossing terms can be ignored.

\section{Integral}
In this section, we wish to evaluate the generic integral
\beq
I(n_+,n_-,n_0|m_a,m_b)=\int_{\mathbb{H}^{d-1}} dY_c\;(-2P_a\cdot P_c)^{-m_a}(-2P_b\cdot P_c)^{-m_b}(E_+\cdot P_c)^{n_+}(E_-\cdot P_c)^{n_-}(E_0\cdot P_c)^{n_0}
\eeq
which appears in the calculation of the modular Hamiltonian term, in the limit where we send the modular Hamiltonian to the Rindler horizon. Using Schwinger parameters, we can rewrite this integral as
\beq
I=\int_0^{\infty}\frac{1}{\Gamma(m_a)}dt_a t_a^{m_a-1}\int_{0}^{\infty}\frac{1}{\Gamma(m_b)}dt_bt_b^{m_b-1}\int_{\mathbb{H}^{d-1}}dY_c\;e^{2(t_a P_a+t_b P_b)\cdot P_c}(E_+\cdot P_c)^{n_+}(E_-\cdot P_c)^{n_-}(E_0\cdot P_c)^{n_0}\nonumber.
\eeq
We will use embedding space coordinates $Y=\left(\frac{1+z^2+(x^i)^2}{2z},\frac{1-z^2-(x^i)^2}{2z},\frac{x^i}{z}\right)$ on $\mathbb{H}^{d-1}$. Analytically continuing the above integral in the $w=e^{i\tau_c}$ plane and sending $\tau_c \to \mp i\infty$, we find
\newcommand{\hY}{\hat{Y}}
\beqn
I&=&\beta_c^{(n_+-n_-)}\int_0^{\infty}\frac{1}{\Gamma(m_a)}dt_a t_a^{m_a-1}\int_{0}^{\infty}\frac{1}{\Gamma(m_b)}dt_bt_b^{m_b-1}\int_{\mathbb{H}^{d-1}}dY_c\;(E_0\cdot P_c)^{n_0}\nonumber\\
&\times & \exp\Big((t_ae^{\mp i\tau_a}+t_be^{\mp i\tau_b})\beta_c+2(t_a Y_a+t_b Y_b)\cdot Y_c+O(\beta_c^{-1})\Big)
\eeqn
where $\beta_c= e^{|\tau_c|}$. Now we partition $n_0$ into two integers $\alpha+\beta=n_0$, and rewrite the above integral as
\beqn
I&=&\frac{\beta_c^{(n_+-n_-)}}{2^{n_0}}\left(E_0\cdot \frac{\pa}{\pa Y_a}\right)^{\alpha}\left(E_0\cdot \frac{\pa}{\pa Y_b}\right)^{\beta}\\
&\times & \int_0^{\infty}\frac{1}{\Gamma(m_a)}dt_a t_a^{m_a-\alpha-1}\int_{0}^{\infty}\frac{1}{\Gamma(m_b)}dt_bt_b^{m_b-\beta-1}\int \frac{d^{d-2}x^idz}{z^{d-1}}\;e^{\beta_c(t_aP_a^{\mp}+t_bP_b^{\mp})-|W|\frac{1+z^2+\vec{x}^2}{z}}\nonumber
\eeqn
where in the last line we have rotated $Y_c$ to align $W = t_a Y_a+t_b Y_b$ with $|W|(1,0,\cdots, 0)$. We have also defined
\beq
P^{\pm}_{a,b} = P_{a,b}\cdot E^{\pm}.
\eeq
Then, with the change of variables $z=z'/|W|$, $x=x'/|W|$, the integral becomes (dropping the primes)
\beqn
I&=&\frac{\beta_c^{(n_+-n_-)}}{2^{n_0}}\left(E_0\cdot \frac{\pa}{\pa Y_a}\right)^{\alpha}\left(E_0\cdot \frac{\pa}{\pa Y_b}\right)^{\beta}\int_0^{\infty}\frac{1}{\Gamma(m_a)}dt_a t_a^{m_a-\alpha-1}\int_{0}^{\infty}\frac{1}{\Gamma(m_b)}dt_bt_b^{m_b-\beta-1}\nonumber\\
&\times &e^{\beta_c(t_aP_a^{\mp}+t_bP_b^{\mp})}\int \frac{d^{d-2}x^idz}{z^{d-1}}\;e^{-\frac{|W|^2}{z}-\frac{z^2+\vec{x}^2}{z}}\nonumber\\
&=&\frac{\beta_c^{(n_+-n_-)}}{2^{n_0}}\left(E_0\cdot \frac{\pa}{\pa Y_a}\right)^{\alpha}\left(E_0\cdot \frac{\pa}{\pa Y_b}\right)^{\beta}\mathcal{I}.
\eeqn
Let us focus on $\mathcal{I}$ for the moment. We now change the order of the $(t_a,t_b)$ and $(z,x^i)$ integrals, and rescale $t_{a,b}=\sqrt{z}t'_{a,b}$
\beq
\mathcal{I}=\int \frac{d^{d-2}x^idz}{z^{d-1}}\;z^{(m_a+m_b-n_0)/2}e^{-\frac{z^2+\vec{x}^2}{z}}
\int_0^{\infty}\frac{1}{\Gamma(m_a)}dt_a t_a^{m_a-\alpha-1}\int_{0}^{\infty}\frac{1}{\Gamma(m_b)}dt_bt_b^{m_b-\beta-1}e^{\sqrt{z}\beta_c(t_aP_a^{\mp}+t_bP_a^{\mp})}e^{W^2}.
\eeq
Finally, performing the $x^i$ integrals, and redefining $z=t_c^2$, we get
\beq
\mathcal{I}=C\int_0^{\infty}\int_0^{\infty}\int_0^{\infty}d^3t\;t_a^{m_a-1}t_b^{m_b-1}t_c^{m_c-1}e^{-\sum_{i,j}t_iA_{ij}t_j}
\eeq
where 
$$m_c=[m_a+m_b-n_0-(d-2)]=(n_++n_-+2),\;\; C = \frac{2\pi^{\frac{d-2}{2}}}{\Gamma(m_a)\Gamma(m_b)}$$
and the matrix $A$ is given by
\beq
A_{ij;\pm} = \left(\begin{matrix}1 & -Y_a\cdot Y_b & \frac{\beta_c}{2} P_a^{\mp}\\ -Y_a\cdot Y_b & 1 & \frac{\beta_c}{2} P_b^{\mp}\\ \frac{\beta_c}{2} P_a^{\mp} & \frac{\beta_c}{2} P_b^{\mp} & 1\end{matrix}\right).
\eeq
Rescaling $t_c \to \beta_c^{-1}t_c$, we obtain 
\beq
\mathcal{I}=\beta_c^{-m_c}C\int_0^{\infty} dt_a dt_b dt_c t_a^{m_a-\alpha-1}t_b^{m_b-\beta-1}t_c^{m_c-1}\exp \Big(-t_a^2-t_b^2-\frac{t_c^2}{\beta_c^{2}}+(t_aP_a^{\mp}+t_bP_b^{\mp})t_c-t_at_b(-2Y_a\cdot Y_b)\Big).
\eeq
We can perform the $t_c$ integral in the limit $\beta_c \to \infty$. The integral converges for $\tau_a, \tau_b$ in a neighborhood of $\pi$, and we can then continue the expression outside this region:
\beq
\mathcal{I}=\beta_c^{-m_c}C\Gamma(m_c)\int_0^{\infty} dt_a dt_b t_a^{m_a-\alpha-1}t_b^{m_b-\beta-1}\frac{1}{\left(-t_aP_a^{\mp}-t_bP_b^{\mp}\right)^{m_c}}\exp \Big(-t_a^2-t_b^2-t_at_b(-2Y_a\cdot Y_b)\Big).
\eeq
Finally switching to new integration variables
\beq
t_a= \sigma\lambda,\;\;\;t_b = \frac{\sigma}{\lambda}
\eeq
and performing the $\sigma$ integration, we obtain
\beq
\mathcal{I}=-\beta_c^{-m_c}C\Gamma(m_c)\Gamma\left(\frac{d-2}{2}\right)\int_0^{\infty}  d\lambda\; \lambda^{m_a-\alpha-m_b+\beta-1}\frac{1}{\left(-\lambda P_a^{\mp}-\lambda^{-1}P_b^{\mp}\right)^{m_c}}\frac{1}{(\lambda^2+\frac{1}{\lambda^2}-2Y_a\cdot Y_b)^{\frac{d-2}{2}}}.
\eeq
So the full integral becomes
\beqn
I(n_+,n_-,n_0|m_a,m_b)&=&-\frac{\beta_c^{-2-2n_-}}{2^{n_0}}C\Gamma(m_c)\Gamma\left(\frac{d-2}{2}\right)\left(E_0\cdot \frac{\pa}{\pa Y_a}\right)^{\alpha}\left(E_0\cdot \frac{\pa}{\pa Y_b}\right)^{\beta}\\
&\times &\int_0^{\infty}  d\lambda\; \lambda^{m_a-\alpha-m_b+\beta-1}\frac{1}{\left(-\lambda P_a^{\mp}-\lambda^{-1}P_b^{\mp}\right)^{n_++n_-+2}}\frac{1}{\Big(-\left(\lambda Y_a+\frac{1}{\lambda}Y_b\right)^{2}\Big)^{\frac{d-2}{2}}}.\nonumber
\eeqn
Using the fact that $E_0\cdot E_0=0$, we can further simplify this to obtain the final expression used in the main text.

\section{Quicker argument for the vanishing of the modular Hamiltonian term}\label{sec:quick}

We would like to give a quick argument that the second line of \eqref{mHterm2}
vanishes. We will do this without passing to the hyperbolic coordinates as in the main
text. We will also not make any attempt to explicitly calculate the modular Hamiltonian integral.
Rather our argument here will be based on scaling symmetry and the operator product expansion
in the CFT. We cut off the integrals over the stress tensor close to the entangling surface
by cutting out a tubular region around $\partial A$ of radius $\delta$ and only integrate
over the remaining region $U_\delta$. Our goal will be to show that this term vanishes as we remove the cutoff
$\delta \rightarrow 0$. This cutoff is related to the cutoff in hyperbolic space used 
in Section~\ref{modHam} via $\delta = 1/\Lambda$. So recall that the term of interest is
\beq
\delta S^{(1)}_{EE} = \frac{1}{2} \int_{U_\delta} d^d \bx_a  \int_{U_\delta} d^d \bx_b  \partial_\mu \zeta_\nu (\bx_a)
 \partial_\lambda \zeta_\sigma (\bx_b)
  \left< \hT^{\mu\nu}(\bx_a)  \hT^{\lambda\sigma}(\bx_b) \widehat{H}_E \right>.
\eeq
Integrating by parts on  $\bx_a$ and $\bx_b$  and using the diffeomorphism and trace ward identities we can rewrite this as:
\beq
\label{mhbterm}
\delta S^{(1)}_{EE} = \frac{1}{2} \int_{\partial U_\delta}  \int_{\partial U_\delta} n^\mu(\bx_a) \zeta^\nu(\bx_a) 
n^\lambda(\bx_b) \zeta^\sigma(\bx_b)  \left< \hT^{\mu\nu} (\bx_a)
\hT^{\lambda \sigma} (\bx_b) \widehat{H}_E \right> + {\rm contact\;terms }.
\eeq
In Appendix C, we have already shown that the contact terms vanish in the limit $\delta =1/\Lambda \rightarrow 0$, and so we focus here on the remaining term coming from the boundary of the tubular region.

Of course had we not cut off the integral around the tubular region then we would be done --- the diffeomorphism Ward identity would leave us   with just the contact terms. 
However we choose to worry about potential divergences around the entangling surface for several reasons. Firstly such terms are generic in entanglement entropy calculations --- although, since we are calculating a finite quantity (the non-local part of the entanglement density), one might expect this to not be an issue. Secondly, when we computed the relative entropy term, an enhancement occured in this region which ruins the naive argument that this term should vanish at least as $\delta^2$; here we are checking that such an enhancement does not occur for the modular Hamiltonian term. 

At this stage it is convenient to rewrite the boundary term in 
\eqref{mhbterm} as a bulk integral \emph{inside} the tubular region $\overline{U_\delta}$
\beq
\delta S^{(1)}_{EE} = \frac{1}{2} \int_{\overline{U_\delta}} d^d \bx_a  \int_{\overline{U_\delta}} d^d \bx_b  \int_0^\infty d x^1 x^1 \int d^{d-2} x^i  \partial_\lambda \zeta_\sigma (\bx_a) \partial_\mu \zeta_\nu (\bx_a)  \left< \hT^{\mu\nu}(\bx_a)
 \hT^{\lambda\sigma}(\bx_b) \hT_{00}(0,x^1,x^i) \right> 
\eeq
where again there is a contact term we have dropped based on the analysis in Appendix C. 
This form is now convenient because we can argue that the leading contribution
as $\delta \rightarrow 0$ is:
\begin{align} \nonumber
\delta S_{EE}^{(1)} & \,\, \mathop{=}^? \,\, \frac{(\pi \delta^2)^2}{2} \int_{\partial A} \int_{\partial A} \partial_\mu \zeta_\nu(x_a^i)
\partial_\lambda \zeta_\sigma(x_b^i) \int_0^\infty dx^1 x^1 \int d^{d-2} x^i \left< \hT^{\mu\nu}(x_a^i) 
\hT^{\lambda\sigma}(x_b^i) \hT^{00}(0,x^1, x^i) 
\right> \\ 
& \qquad \qquad + \ldots + {\rm contact\;terms'}
\end{align}
where we have integrated over the tubular region assuming the relevant integrated three-point function is a constant over this region.  If the term multiplying $(\pi \delta^2)^2$ above can be
shown to be finite as $\delta \rightarrow 0$ then this assumption is true and further we can argue there are no enhancements from this contribution to the modular Hamiltonian term. Unfortunately this is not quite correct; instead, we will use the OPE of two stress tensors to show that there is at most a logarithmic divergence coming from the $x^1$ integral which we should then cut off at small $x^1 \approx \delta$
close to $\partial A$.  This should only lead to a mild enhancement such that the overall
scaling of this term is $\delta^4 \ln \delta$. 
\myfig{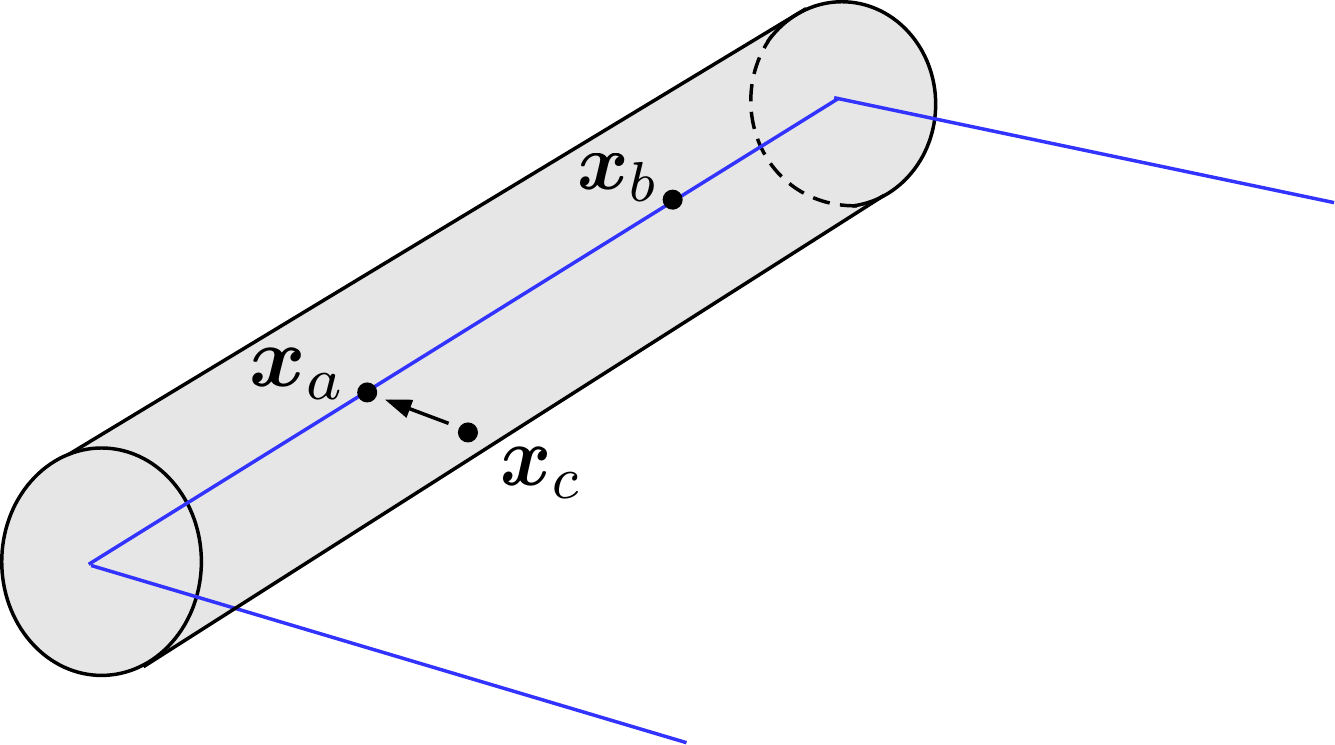}{7.5}{\textsf{The mild logarithmic enhancement comes from the stress tensor in the modular Hamiltonian coming close to one of the other two stress tensor insertions inside $\overline{U_{\delta}}$.}}

The issue comes about for $ x^i \rightarrow x^i_{a,b}$ and $x^1 \rightarrow 0$
where the stress tensor in the modular Hamiltonian comes close to either
one of the two other stress tensor insertions (recall that $x_a$ is well separated from $x_b$
so we need never consider these two operators colliding) -- see Fig.~\ref{fig:fig5}. In this limit we can use the OPE \cite{Osborn:1993cr}:
\beq
\hT^{00}(0,x^1,x^i) \hT^{\mu\nu} ( x_a^i) \rightarrow |\sigma|^{-d} A^{00;\mu\nu;\alpha \beta} (\hat{\sigma})
\hT_{\alpha\beta}(x^i_a) 
+ \ldots
\eeq
where $\sigma= (0,x^1,x^i-x_a^i)$ and $\hat{\sigma} = \sigma/|\sigma|$. The function $A$ depends on several
conformally covariant structures with three unfixed theory-dependent parameters. These are
the same parameters that appear in the stress tensor three-point function.

Plugging this into the three-point function we find the leading behavior:
\beq
\delta S_{EE}^{(1)} =\frac{(\pi \delta^2)^2}{2} \int_{\partial A} \int_{\partial A}
 C^{00;\mu\nu;\alpha\beta}   \partial_\lambda \zeta_\sigma (x_a^i) \partial_\mu \zeta_\nu (x_a^i)   \left<\hT_{\alpha\beta}(x^i_a) \hT^{\lambda\sigma}(x_b^i)   \right> + \ldots
\eeq
where:
\beq
\label{diva}
C^{00;\mu\nu;\alpha\beta} = \left. \int_\delta d x^1 x^1 \int d^{d-2} y^i |\sigma|^{-d} A^{00;\mu\nu;\alpha\beta}
(\hat{\sigma}) \right|_{\sigma = (0,x^1,y^i)} 
\eeq
This integral is log divergent close to $\sigma=0$ which we cut off by hand at $x^1 \sim \delta$.  We justify this since  for $x^1 \sim \delta$ we cannot replace the integral over the tubular region by an integral over $\partial A$.
Then we can argue that $C^{00;\mu\nu;\alpha\beta} \sim \alpha \ln (\delta/|x^i_a -  x^j_b|) + \beta$, where the $\log$ is cut off at long distances when the OPE expansion breaks down. This argument
does not fix $\beta$. It is also possible to show using the same OPE argument that the $\log$ is indeed
the only enhancement possible for $x^1 \sim \delta$ and this argument also leads to an explicit and finite expression for $\beta$. 
We leave this as an exercise to the reader.

We note that the results of this Appendix suggest that this contribution to the entanglement density
scales as $\delta^4 \ln \delta$, however in the main text  we gave an argument that this term
should vanish at least as $\delta^2 = \Lambda^{-2}$ as suggested in \eqref{expl}. For consistency with the results in this Appendix the $\mathcal{O}(\Lambda^{-2})$ term in \eqref{expl} should actually vanish after we integrate that term over $\tau_a$ and $\tau_b$ in \eqref{3pt1} and the leading term should come in at $\mathcal{O}(\Lambda^{-4})$ with a possible logarithmic enhancement from the $\lambda$ integral in \eqref{lamint}. 
We have checked that this is indeed the case. 

\providecommand{\href}[2]{#2}\begingroup\raggedright\endgroup


\end{document}